\documentclass[twocolumn,prd,aps,showpacs]{revtex4}

\usepackage{verbatim}
\usepackage{graphicx}
\usepackage[usenames]{color}
\usepackage{psfrag}
\usepackage[ps2pdf,colorlinks,bookmarks]{hyperref}

\definecolor{linkblue}{rgb}{0,0,0.8}
\definecolor{linkgreen}{rgb}{0,0.5,0}
\hypersetup{pdfpagemode=None, pdfstartview=FitH, linkcolor=linkblue, %
            citecolor=linkgreen, urlcolor=linkblue}

\newcommand{\ud}[2]{^{#1}_{\phantom{#1}#2}}

\def\beq{\begin{equation}}
\def\eeq{\end{equation}}
\def\bea{\setlength\arraycolsep{1.4pt}\begin{eqnarray}}
\def\eea{\end{eqnarray}}
\def\bit{\begin{itemize}}
\def\eit{\end{itemize}}
\def\nn{\nonumber}

\def\ie{{i.e.}}
\def\eg{{e.g.}}

\def\eq{Eq.~}
\def\eqs{Eqs.~}
\def\fig{Fig.~}

\def\pd{\partial}
\def\lap{\nabla^2}
\def\lapp{\fr{1}{a^2}\nabla^2}

\def\ld{\left}
\def\rd{\right}
\def\ra{\rightarrow}
\def\tl{\tilde}

\def\ph{\phantom}
\def\fr{\frac}
\def\oo{\frac{1}}
\def\half{\frac{1}{2}}

\def\gam{\gamma}

\def\del{\delta}
\def\kap{\kappa}
\def\lam{\lambda}
\def\Lam{\Lambda}

\def\sig{\sigma}
\def\Sig{\Sigma}
\def\om{\omega}
\def\Om{\Omega}

\def\Rs{{}^{(3)}\!R}

\def\O{{\cal O}}

\def\R{{\cal R}}

\def\gtrn{gauge transformation}

\def\hsf{hypersurface}
\def\hsfs{hypersurfaces}
\def\pert{perturbation}
\def\perts{perturbations}

\def\emt{energy-momentum tensor}

\def\bra{\langle}
\def\ket{\rangle}

\def\PR{{\cal P_{\cal R}}}
\def\Rpr{\mathcal{R}^{\rm pr}}
\def\rls{r_{\rm LS}}

\def\etals{\eta_{\rm LS}}
\def\jl{j_\ell}
\def\Hb{\bar H}
\def\Sls{\Sigma_{\rm LS}}

\begin{document}

\title{Gauging the cosmic microwave background}

\author{J. P. Zibin} \email{zibin@phas.ubc.ca}
\author{Douglas Scott} \email{dscott@phas.ubc.ca}
\affiliation{Department of Physics \& Astronomy, %
University of British Columbia, %
Vancouver, BC, V6T 1Z1  Canada}

\date{\today}

\begin{abstract}
   We provide a new derivation of the anisotropies of the cosmic microwave 
background (CMB), and find an exact expression that can be readily expanded 
perturbatively.  Close attention is paid to gauge issues, with the 
motivation to examine the effect of super-Hubble modes on the CMB.  We 
calculate a transfer function that encodes the behaviour of the dipole, 
and examine its long-wavelength behaviour.  We show that contributions 
to the dipole from adiabatic super-Hubble modes are strongly suppressed, 
even in the presence of a cosmological constant, contrary to claims in the 
literature.  We also introduce a naturally defined CMB monopole, which 
exhibits closely analogous long-wavelength behaviour.  We discuss the 
geometrical origin of this super-Hubble suppression, pointing out that 
it is a simple reflection of adiabaticity, and hence argue that it will 
occur regardless of the matter content.
\end{abstract}


\maketitle


\section{Introduction}

   The anisotropies in the cosmic microwave background (CMB) reveal a 
great deal about our Universe, since they persist essentially unscathed 
from the epoch when fluctuations were well described by simple linear 
theory.  The comparison of CMB observations with theory has become a 
mature subject, and has played an important role in forming our current 
understanding of the Universe (see, \eg, \cite{wmap5}).

   Critical to that comparison is the accurate theoretical calculation 
of the anisotropies.  Since the pioneering work of Sachs and Wolfe 
\cite{sw67}, the theoretical anisotropies have been refined and 
recalculated using different formalisms many times (see, \eg, 
\cite{panek86,magueijo92,rsxd93,wh97,dunsby97,cl98,hn99}).  Accurate 
calculations are now readily available via public code packages such as 
\textsc{camb} \cite{lcl00,notecamb}.

   In the present work, we revisit the calculation of anisotropies.  
While our results may not lead to more accurate or efficient 
calculations, we hope that they will help to clarify some of the 
conceptual issues surrounding the calculations.  In particular, our 
approach makes explicit the physical meaning of the various 
contributions to the anisotropies.  Crucial to this is our use of 
the covariant approach to cosmology (see \cite{ee98,tcm08} for reviews), 
which is ideal for writing exact solutions and for physical clarity.  
We present a remarkably simple but exact expression for 
the anisotropy, which applies to arbitrary spacetimes and includes the 
effects of tensor as well as scalar \perts\ and any line-of-sight 
integrated Sachs-Wolfe (ISW) effect.  This general result can be 
readily expanded perturbatively, and here we turn to the metric 
formalism for computational efficiency and show that we recover 
previous results in the literature.

   A main motivation for our work is in examining the behaviour of 
the anisotropies due to super-Hubble fluctuations (we use the terms 
``super-Hubble'', ``long-wavelength'', and ``large-scale'' 
interchangeably, to mean scales larger than the current Hubble 
or last scattering radius), where gauge issues are paramount.  This 
question has been examined before 
in the context of the Grishchuck-Zel'dovich effect 
\cite{gz78}, which describes the large-angular-scale anisotropies 
that result from super-Hubble modes.  In the context of a matter-dominated 
universe with adiabatic fluctuations, it was shown that the CMB 
dipole receives strongly suppressed contributions from long-wavelength 
modes.  A claim was made in Ref.~\cite{turner91} that this suppression 
would not occur in models with cosmological constant, so that we 
could ``see'' very long-wavelength structure in the dipole.  It was 
also found that in the presence of {\em isocurvature} \perts, the 
suppression may not occur (see \cite{langlois96} and references 
therein).

   To study this issue, we construct a transfer function that describes 
the scale dependence of contributions to the dipole.  Working by 
analogy, we carefully define a CMB {\em monopole} \pert, and find 
its transfer function.  As is well known, such a monopole cannot be 
observable, but we show that its {\em variance} is well defined 
theoretically.  Our definitions have simple interpretations:  the 
dipole measures the departure of radiation and matter comoving 
worldlines, while the monopole measures how well radiation and matter 
constant-density hypersurfaces coincide.  The usefulness of the monopole 
will be in examining its long-wavelength behaviour, where it will 
help to clarify the dipole case.

   We show that the contributions to both dipole and monopole vanish 
for large scale sources, even in the presence of a cosmological 
constant.  We close by pointing out that this is a direct consequence 
of adiabaticity, and hence that this result is expected to hold regardless 
of the matter content of the Universe, unless isocurvature modes are 
present.

   Another potential reason that a careful treatment of the monopole 
may be of interest involves the measurement of the mean CMB temperature 
and its relation to constraints on other cosmological parameters.  
The mean temperature is currently measured to a precision of a few 
parts in $10^4$ \cite{mfsmw99}.  It has been pointed out that the 
precision of this measurement could be improved by nearly two orders 
of magnitude with currently available technology \cite{fm02}.  Such 
a measurement would reach the naive cosmic variance limit of a part 
in $10^5$ as suggested by the observed amplitude of fluctuations 
(see \cite{wz08} for a related discussion).  It would then become 
necessary to be very careful about exactly what information the mean 
temperature measurement is giving us, and about the nature of monopole 
fluctuations.  Relevantly, recent studies have examined the importance 
of the mean temperature measurement to our ability to constrain the 
cosmological parameters \cite{cs08,hw08}.

   Since the covariant approach to cosmology is essential to this work, 
we begin in Sec.~\ref{covarsec} with a summary of the required formalism.  
Next, in Sec.~\ref{swsec} we present the derivation of the Sachs-Wolfe 
effect, beginning with an exact result before specializing to first 
order and recovering previous results.  In Sec.~\ref{secmonodi} we 
present calculations of the dipole and monopole transfer functions, and 
we examine their long-wavelength behaviour in Sec.~\ref{seclongwl}.  
Finally we discuss our results in Sec.~\ref{discusssec}.  The Appendices 
summarize relevant material in the metric formalism, and demonstrate 
both the gauge invariance and the gauge dependence of our results.  
We use signature $(-,+,+,+)$, and greek indices indicate four-tensors, 
while latin indices indicate spatial three-tensors.

\section{Covariant cosmology}
\label{covarsec}

   This section will contain a brief summary of the elements of the 
covariant approach to cosmology (see, \eg, the reviews \cite{ee98,tcm08}) 
that will be needed in the following sections.  (A collection of results 
from the metric-based approach to cosmological \perts, which will 
also be needed, is presented in Appendix~\ref{secmetappr}.)  Fundamental to 
the covariant approach is the notion of a {\em congruence} of worldlines, 
also known as a {\em threading} of the spacetime or sometimes as a choice of 
{\em frame}.  This is a family of worldlines such that exactly one 
worldline passes through each event.  A useful example is the congruence 
of comoving worldlines, when it is well defined.  A timelike congruence is 
described by a vector 
field $u^\mu$, tangent everywhere to the worldlines.  We will assume that 
$u^\mu$ is normalized, $u^\mu u_\mu = -1$.  At each event we can define the 
spatial projection tensor
\beq
h\ud{\mu}{\nu} \equiv \del\ud{\mu}{\nu} + u^\mu u_\nu,
\label{spatmetdef}
\eeq
which projects orthogonal to $u^\mu$.  Hypersurfaces orthogonal everywhere 
to $u^\mu$ will exist if the twist of the congruence (defined below) 
vanishes \cite{wald84}, in which case $h_{\mu\nu}$ is the (Riemannian) metric 
tensor for those spatial \hsfs.

   It is useful to describe the geometrical properties of the congruence 
by the covariant derivative $u_{\mu;\nu}$.  By virtue of the normalization 
condition, this derivative satisfies
\beq
u_{\mu;\nu}u^\mu = 0.
\label{uspatial}
\eeq
The derivative can be decomposed into parts parallel to and orthogonal 
to $u^\mu$ in its second index using \eq(\ref{spatmetdef}), giving
\beq
u_{\mu;\nu} = h\ud{\rho}{\nu}u^{}_{\mu;\rho} - u^\rho u_\nu u_{\mu;\rho}.
\label{ucovderst}
\eeq
The temporal part can be written in terms of the acceleration of the 
worldlines, defined by
\beq
a_\mu \equiv u_{\mu;\rho}u^\rho.
\label{acceldef}
\eeq
The field $a^\mu$ measures the departure of the worldlines from 
geodesic.  The spatial part 
$h\ud{\rho}{\nu}u^{}_{\mu;\rho}$ can be decomposed into trace, 
symmetric trace-free, and antisymmetric parts via
\bea
\theta        &\equiv& h^{\rho\mu}u^{\ph{\rho}}_{\mu;\rho} = u\ud{\mu}{;\mu},\\
\sig_{\mu\nu} &\equiv& h\ud{\rho}{(\nu}u^{\ph{\rho}}_{\mu);\rho}
                     - \oo{3}\theta h_{\mu\nu},\label{sigdef}\\
\om_{\mu\nu}  &\equiv& h\ud{\rho}{[\nu}u^{\ph{\rho}}_{\mu];\rho}.
\label{twistdef}
\eea
The scalar $\theta$ and tensors $\sig_{\mu\nu}$ and $\om_{\mu\nu}$ 
measure the local rates of expansion, shear, and twist of the 
congruence, respectively.  Combining \eqs(\ref{ucovderst}) to 
(\ref{twistdef}) we have
\beq
u_{\mu;\nu} = \oo{3}\theta h_{\mu\nu} + \sig_{\mu\nu} + \om_{\mu\nu}
            - a_\mu u_\nu.
\label{ucovder}
\eeq
For the case of a homogeneous and isotropic Friedmann-Robertson-Walker (FRW) 
cosmology, if we choose $u^\mu$ to be comoving then $\sig_{\mu\nu} = 
\om_{\mu\nu} = a_\mu = 0$ and $\theta = 3H$, where $H \equiv \dot a/a$ is 
the Hubble rate, and $a$ the scale factor.

   We will also need two derivatives.  For an arbitrary tensor $X$, the 
covariant time derivative is defined by $\dot X \equiv X_{;\mu}u^\mu$, and 
gives the proper time derivative along $u^\mu$.  The symbol $D_\mu$ 
represents the spatial (orthogonal to $u^\mu$) covariant derivative 
defined using the spatial metric $h_{\mu\nu}$.  For example,
\beq
D^{}_\nu X^\mu \equiv h\ud{\lam}{\nu} h\ud{\mu}{\rho} X\ud{\rho}{;\lam},
\label{spatcovder}
\eeq
for any tensor $X^\mu$ orthogonal to $u^\mu$.

   A congruence $u^\mu$ can be used to describe the matter content as 
observed locally by a family of observers, given the \emt\ $T_{\mu\nu}$.  
The energy density $\rho$, momentum density $q_\mu$, pressure $P$, and 
anisotropic (shear) stress $\pi_{\mu\nu}$, as viewed locally by an 
observer with four-velocity $u^\mu$, are defined by the projections
\bea
T_{\mu\nu}u^\mu u^\nu                       &=& \rho,\\
T^{}_{\mu\nu}u^\mu h\ud{\nu}{\kap}          &=& -q_\kap,\\
T^{}_{\mu\nu}h\ud{\mu}{\lam}h\ud{\nu}{\kap} &=& Ph_{\lam\kap} + \pi_{\lam\kap},
\eea
where $\pi_{\mu\nu}$ is defined to be trace-free.

\section{Sachs-Wolfe effect}
\label{swsec}

   In this section, we will provide a calculation of the anisotropies of 
the CMB radiation.  The main goal will be to clarify the issues of 
gauge and frame choice, which will be critical to properly describing the 
dipole and monopole anisotropies in the next section.  For this reason, 
it will not be necessary to consider here the finite thickness of the 
last scattering surface, or the effect of reionization on the 
anisotropies, which have negligible effect on the largest-scale 
anisotropies.  Thus we assume tight coupling in the baryon-photon plasma 
before last scattering, followed by instantaneous recombination of matter 
and free streaming (\ie\ unattenuated geodesic evolution) of the radiation.  
Apart from this approximation the calculation will be exact.

\subsection{Exact expression}

   We are making the approximation that the CMB radiation is emitted 
abruptly when the local plasma temperature drops below some value $T_E$ at 
which recombination occurs ($E$ for ``emission''), and thereafter travels 
freely.  This temperature is defined with respect to the frame (or 
congruence) $u^\mu$ which is comoving with the plasma (\ie\ for which 
the plasma momentum density vanishes).  The spacelike hypersurface 
defined by the moment of recombination will be 
called the last scattering hypersurface, $\Sls$, while the intersection 
of $\Sls$ with an observer's past light cone defines the two-sphere 
commonly called the observer's last scattering surface (LSS).  Via the 
Stefan-Boltzmann law, $\Sls$ must be a \hsf\ of constant radiation 
energy density $\rho_{(\gam)}$, with the density again defined with 
respect to the comoving congruence $u^\mu$ \cite{noterhoboost}.  
For adiabatic 
perturbations at last scattering and on large scales, $\Sls$ will also be 
at constant matter energy density, and hence total energy density $\rho$.  
The definition of the last scattering \hsf\ is critical here:  note, in 
particular, that it does {\em not} contain colder and hotter regions which 
contribute to the anisotropies observed at later times, as some accounts 
state (see, \eg, \cite{wss94,wh97,bmr05}) \cite{notebgndtemp}.  
Rather, it is a \hsf\ of constant (comoving) temperature.  
Nevertheless, in a realistic cosmology, $\Sls$ is still a perturbed 
surface, with generally non-vanishing intrinsic and extrinsic curvature 
and matter perturbations apart from $\rho_{(\gam)}$, and these \perts\ 
will source anisotropies.  Also, because of these \perts, the comoving 
congruence will not in general be orthogonal to $\Sls$.

\begin{figure}[ht]\begin{center}
\includegraphics[width=\columnwidth]{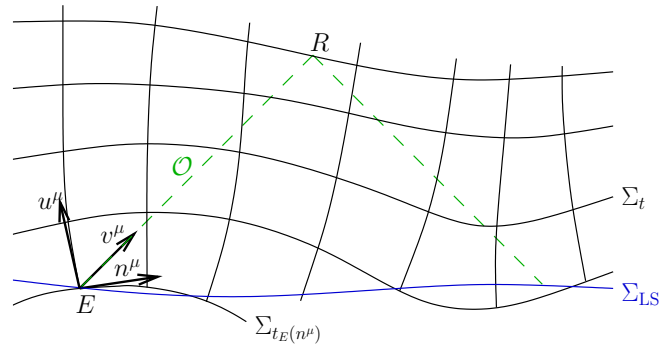}
\caption{A conformal spacetime diagram showing a light ray $\O$ emitted from 
the LSS on $\Sls$ at event $E$ and received at $R$.  A foliation $\Sig_t$ is 
indicated together with its orthogonal congruence $u^\mu$, which is comoving 
on the LSS and at $R$.  For clarity, orthogonal vectors are displayed as 
if the geometry was Euclidean.}
\label{fig1}
\end{center}\end{figure}

   In order to calculate the anisotropies observed at late times we 
must propagate the radiation along null geodesics from the observer's LSS.  
Consider a light ray following a null geodesic $\O$ with tangent vector 
$v^\mu$ that extends from a 
point of emission, $E$, on the observer's LSS to a point of reception, $R$, 
and also consider a timelike congruence $u^\mu$ defined in the vicinity of 
$\O$.  We define the congruence in order to provide a frame with respect 
to which a local energy density (and hence temperature) can be expressed.  
Then we can decompose $v^\mu$ at each point on $\O$ into parts parallel and 
orthogonal to $u^\mu$ according to
\beq
v^\mu = \gamma (u^\mu + n^\mu),
\label{vdecomp}
\eeq
with $n^\mu u_\mu = 0$ and $n^\mu n_\mu = 1$.  The spatial vector $n^\mu$ 
defines the spatial direction of propagation of the light ray at each 
point along $\O$.  Figure~\ref{fig1} illustrates this geometry.

   Since the radiation is 
emitted from $E$ with a thermal spectrum, an observer at any point on 
$\O$, with four-velocity $u^\mu$, will observe the radiation travelling 
along $\O$ to also have a thermal spectrum with temperature
\beq
T \propto -u^\mu v_\mu = \gamma.
\label{tproptogamma}
\eeq
Because the temperature at emission, $T_E$, is defined with respect to the 
plasma frame and is constant on $\Sls$, the calculation will be 
simplest if we choose $u^\mu$ to be comoving with the plasma on the 
observer's LSS.  Similarly, it will 
be most natural to choose $u^\mu$ to be comoving at $R$, since any observer 
will presumably be composed of matter and comoving with it.  Note that such an 
observer, comoving with the total (effectively matter) energy density at $R$, 
will generally not be comoving with the radiation, and so will observe a 
dipole anisotropy.  As we will see, the congruence can be freely chosen in 
between $E$ and $R$, although some choices will be computationally more 
efficient in practice.  In Sec.~\ref{seclinarbgsw} we will relax the 
constraint that $u^\mu$ be comoving at $E$ and $R$.

   Now we will derive an expression for the evolution of the ``redshift 
parameter'' $\gamma$ along $\O$, which will tell us how the observed 
temperature evolves, using only the geodesic equation,
\beq
v\ud{\mu}{;\nu}v^\nu = 0.
\label{vgeod}
\eeq
Consider the quantity
\bea
H_n \equiv \oo{\gam^2}u_{\mu;\nu}v^\mu v^\nu &=&
    u_{\mu;\nu}n^\mu n^\nu + a_\mu n^\mu\\
&=& \oo{3}\theta + \sig_{\mu\nu}n^\mu n^\nu + a_\mu n^\mu,\label{Hnexact}
\eea
which is related to the expansion rate of the congruence $u^\mu$ 
projected into the plane defined by $u^\mu$ and $n^\mu$ \cite{noteextrcurv} 
(indeed $H_n$ 
reduces to the familiar Hubble rate for the comoving congruence in a 
homogeneous and isotropic spacetime).  Now, using \eqs(\ref{tproptogamma}) 
and (\ref{vgeod}), we have
\bea
H_n &=& \oo{\gam^2}(u_\mu v^\mu)_{;\nu} v^\nu\\
    &=& \fr{d(\gam^{-1})}{d\lam},\label{hgamma}
\eea
where $\lam$ is an affine parameter along $\O$ (note that this last 
expression is independent of the affine parameter chosen).  The expression 
\eq(\ref{hgamma}) gives the exact evolution of the redshift 
parameter $\gam$ along 
$\O$ with respect to the congruence $u^\mu$ in an arbitrary spacetime.  It 
describes the familiar behaviour of increasing redshift (decreasing $\gam$) 
in an expanding universe, with ``expansion'' now seen to mean precisely 
that $H_n > 0$.

   We can in principle integrate \eq(\ref{hgamma}) along $\O$ to find the 
temperature $T_R(n^\mu)$ observed at $R$ in direction $-n^\mu$, \ie\ the 
CMB temperature sky map.  With an appropriate choice of affine parameter 
[setting the proportionality constant equal to unity in 
\eq(\ref{tproptogamma})], we have \cite{notelensiso}
\beq
T_R(n^\mu) = \ld(T_E^{-1} + \int_{E(n^\mu)}^R H_n d\lam\rd)^{-1}.
\label{tnaffine}
\eeq
However, in practice it would be very difficult to determine $H_n$ as 
a function of affine parameter along each null geodesic in order to 
perform the integral in \eq(\ref{tnaffine}) for a particular spacetime.  
Instead, if we define a time coordinate $t$ by foliating the spacetime 
into spacelike \hsfs\ of constant $t$, $\Sig_t$, (see \fig\ref{fig1}) 
we can transform the integral into one that is more tractable in terms 
of the new coordinate.  The most convenient choice for the foliation 
is that which is everywhere orthogonal to the congruence $u^\mu$ 
\cite{notetwist}.

   Using \eq(\ref{vdecomp}), we can write
\bea
\ld.\fr{d(\gam^{-1})}{d\lam}\rd|_{v^\mu}
   &=& \gam\ld.\fr{d(\gam^{-1})}{d\tau}\rd|_{u^\mu}\\
   &=& \fr{\gam}{N}\ld.\fr{d(\gam^{-1})}{dt}\rd|_{u^\mu}\\
   &=& \fr{\gam}{N}\ld.\fr{d(\gam^{-1})}{dt}\rd|_{v^\mu}.\label{gammat}
\eea
Here $\tau$ is proper time, $N$ is the lapse function for the slicing 
$\Sig_t$, and the subscript $v^\mu$ or $u^\mu$ indicates the direction in 
which the derivative is taken.  In the intermediate steps we have chosen 
to define $\gam$ away from $\O$ to be constant along the $\Sig_t$.  
Integrating \eq(\ref{gammat}) along $\O$ and using \eq(\ref{hgamma}) and the 
proportionality \eq(\ref{tproptogamma}) between $\gamma$ and temperature, 
we finally obtain
\beq
\fr{T_R(n^\mu)}{T_E} = \exp\ld({\int_{t_R}^{t_E(n^\mu)} H_n N dt}\rd),
\label{swexact}
\eeq
where $t_E(n^\mu)$ is the value of $t$ at the point of emission $E$ on the 
LSS corresponding to the observed direction $-n^\mu$ at $R$, and $t_R$ is the 
time of observation.  [Note that in general $\Sig_{\rm LS}$ will not coincide 
with one of the slices $\Sig_t$; hence the dependence $t_E(n^\mu)$.]  This 
remarkably simple expression is exact, and so is not restricted to linear, 
adiabatic, or scalar fluctuations, and it applies to all scales (subject of 
course to our basic assumption of abrupt recombination).  In particular, 
\eq(\ref{swexact}) encapsulates in principle the acoustic peak structure 
of the CMB, any ISW 
contribution that may arise, as well as the effect of gravitational waves.  
This equation is a purely geometrical result, independent of any 
dynamical input such as stress-energy conservation or Einstein's equations.  
To our knowledge \eq(\ref{swexact}) has not been written down before, 
although a related expression appears in Ref.~\cite{dunsby97}, and a 
related linearized expression appears in Ref.~\cite{sxek95}.

   Equation (\ref{swexact}) tells us that the observed temperature in some 
direction on the sky is determined entirely by the integrated line-of-sight 
component of expansion (or 
number of ``$e$-folds'') along the null path from the LSS, of a congruence 
that is comoving with the plasma at the LSS and matches the observer's 
four-velocity at $R$.  Thus we can interpret the observed anisotropic CMB 
sky as a uniform temperature surface viewed through an anisotropically 
expanding universe:  the observed hot and cold spots on the sky are simply 
``closer'' and ``farther'', respectively, from us, in terms of $e$-folds 
of expansion.  Note, however, that we cannot view the redshifting as 
uniquely defined at intermediate points between $E$ and $R$, since we are 
free to deform the congruence between those endpoints.  Rather, it 
is only the total integral that is independent of the choice of congruence 
$u^\mu$ between the endpoints, when we fix the position and state of 
motion of the observer at $R$.  For the special case of a homogeneous 
and isotropic FRW cosmology, and choosing the comoving congruence, for 
which $H_n = H$ and $N = 1$, we immediately recover from 
\eq(\ref{swexact}) the familiar result for the cosmological redshift
\beq
\fr{T_R}{T_E} = \exp\ld({\int_{t_R}^{t_E} H dt}\rd) = \fr{a_E}{a_R},
\eeq
for scale factor $a$.

\subsection{Linearized results}
\subsubsection{Arbitrary gauge}
\label{seclinarbgsw}

   The result \eq(\ref{swexact}) is very general, but probably has limited 
direct use for calculating anisotropies.  However, it can be 
straightforwardly expanded to linear (or even higher) order in perturbation 
theory, and such a linearized calculation will be very convenient, as 
linear theory captures very well the evolution of structure at early times 
and very large scales today.  In doing this it will prove 
helpful to generalize the result to congruences $u^\mu$ that are non-comoving 
at $E$ and $R$.  With such threadings it will then be necessary to provide 
explicit boosts at the LSS and at the reception point $R$ to compensate.  
Similarly, at linear order it will be simple to write the integral to the LSS 
in terms of an integral to some constant time slice, plus a contribution 
due to the linear temporal 
displacement to the actual LSS.  The boost at the LSS will constitute what is 
often termed the ``Doppler'' or ``dipole'' contribution to the anisotropies, 
while the temporal displacement contributes to what is sometimes called the 
``monopole'' contribution.

\begin{figure}[ht]\begin{center}
\includegraphics[width=\columnwidth]{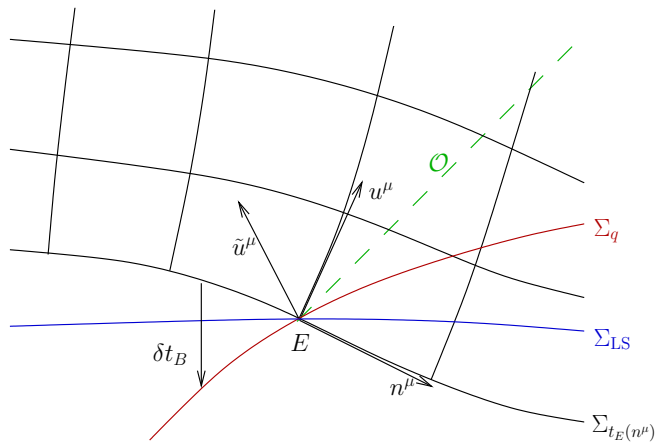}
\caption{A conformal spacetime diagram showing a light ray $\O$ emitted from 
the LSS at $E$.  The arbitrary foliation $\Sig_t$ is indicated together 
with its orthogonal congruence $u^\mu$.  Also, the \hsf\ $\Sig_q$ 
orthogonal to the direction $\tl u^\mu$ comoving with the plasma is 
indicated, together with the ``boost displacement'' $\del t_B$, which 
takes $\Sig_{t_E(n^\mu)}$ into $\Sig_q$.}
\label{fig2}
\end{center}\end{figure}

   To calculate the boosts, consider the general non-comoving congruence 
$u^\mu$ and the direction $\tl u^\mu$ comoving with the plasma at 
emission point $E$ (see 
\fig\ref{fig2}).  If we temporarily construct scalar fields $t$ and 
$\tl t$ in the vicinity of $E$ such that $u^\mu = -t^{;\mu}$ and $\tl u^\mu 
= -\tl t^{;\mu}$, with $\tl t \equiv t - \delta t_B$, then we have
\beq
\tl u^\mu = u^\mu + \delta t_B^{;\mu}.
\eeq
While this expression is exact, the ``boost displacement'' $\delta t_B$ 
evaluated at linear order is simply the linear temporal displacement 
between \hsfs\ orthogonal to $u^\mu$ and $\tl u^\mu$, in the vicinity of 
$E$ (see \fig\ref{fig2}); this can be readily calculated in the metric 
formalism as the gauge 
transformation required to take the gauge specified by the slicing 
orthogonal to $u^\mu$ into the plasma-comoving gauge.  To calculate the 
change in observed temperature due to the boost, we require the quantity
\beq
\tl\gam \equiv -\tl u^\mu v_\mu = \gam\ld(1 - n_\mu \del t_B^{;\mu}\rd),
\label{boost}
\eeq
which is valid at first order.  To derive this expression we have used the 
fact that $u_\mu \del t_B^{;\mu}$ vanishes as first order, which follows 
from the normalization 
of the four-velocities.  Note that \eq(\ref{boost}) simply describes a 
local Lorentz transformation of photon energy.  This expression can also 
be applied at the reception point $R$, although the direction $\tl u^\mu$ 
is free in principle there.  If we choose the observer to be comoving 
with matter then the displacement $\delta t_B$ at $R$ will be given by 
the gauge transformation required to take the gauge specified by $u^\mu$ 
into the comoving gauge.  The freedom to choose $\tl u^\mu$ at $R$ only 
effects the dipole anisotropy at linear order, as shown in 
Appendix~\ref{gdepapp}.

   To calculate the temporal displacements at $E$ and $R$, we can write the 
integral in the exact expression \eq(\ref{swexact}) as
\beq
\int_{t_R}^{t_E(n^\mu)} H_n N dt
   = \int_{\bar{t}_R + \del t_D(R)}^{\bar{t}_E + \del t_D(E)} H_n N dt,
\eeq
where $\bar{t}_E$ and $\bar{t}_R$ label particular slices $\Sig_t$, and can 
be considered the background emission and reception times.  The displacement 
$\del t_D(E)$ accounts for the separation between the background slice 
$\Sig_{\bar{t}_E}$ and the true last scattering \hsf\ $\Sig_{\rm LS}$, 
and is a function of $n^\mu$.  Similarly, the displacement $\del t_D(R)$ 
accounts for the separation between the background slice 
$\Sig_{\bar{t}_R}$ and the actual slice on which the reception point $R$ 
is located.  $\del t_D(R)$ can be considered a function of position 
if we wish to evaluate the anisotropies at various reception points $R$.  
At linear order, we can then write
\beq
\int_{t_R}^{t_E(n^\mu)} H_n N dt = \int_{\bar{t}_R}^{\bar{t}_E} H_n N dt
   + \ld.\ld(\bar{H}\del t_D\rd)\rd|^E_R,
\label{tdispl}
\eeq
where $\bar{H}(t)$ is the zeroth order (background) Hubble rate and we have 
assumed that the lapse equals unity at zeroth order, so that the 
coordinate $t$ is a perturbed proper time.

   The displacement $\del t_D(E)$, like the boost displacements, can be 
readily calculated in the metric formalism as the gauge transformation 
required to take the gauge specified by the slices $\Sig_t$ (orthogonal to 
$u^\mu$) into the gauge specified by $\Sig_{\rm LS}$, \ie\ the uniform 
radiation energy density gauge.  The displacement $\del t_D(R)$ is not 
fixed uniquely by any such physical prescription, but only affects the 
monopole anisotropy, as shown in Appendix~\ref{gdepapp}.

   The temperature observed at $R$ in direction $-n^\mu$ can now be written 
using the boost relation \eq(\ref{boost}) as
\beq
\fr{T_R(n^\mu)}{T_E} = \fr{\tl\gam_R(n^\mu)}{\tl\gam_E}
                     = \fr{\gam_R(n^\mu)}{\gam_E(n^\mu)}
                       \ld(1 - n_\mu\del t_B^{;\mu}\Big|_E^R\rd),
\label{tempboost}
\eeq
at linear order.  The prefactor on the right-hand side of this equation is 
simply the redshift due to the expansion along the congruence $u^\mu$, so 
it is given by \eq(\ref{swexact}).  Writing
\beq
H_n N = \bar{H} + \del(H_n N)
\label{hdecomp}
\eeq
and using \eq(\ref{tdispl}), we have
\bea
\fr{\gam_R(n^\mu)}{\gam_E(n^\mu)}
  &=& \exp\ld({\int_{t_R}^{t_E(n^\mu)} H_n N dt}\rd)\\
  &=& \fr{\bar{T}_R}{T_E} 
      \ld(1 + {\int_{\bar{t}_R}^{\bar{t}_E} \del(H_nN)dt}
   + \ld.\ld(\bar{H}\del t_D\rd)\rd|^E_R\rd)\quad
\label{tempnoboost}
\eea
at linear order, where we have defined the ``background'' observed temperature 
$\bar{T}_R$ by
\beq
\bar{T}_R \equiv T_E\exp\ld({\int_{\bar{t}_R}^{\bar{t}_E} \bar{H}dt}\rd).
\label{tempbgnd}
\eeq
Finally, combining \eqs(\ref{tempnoboost}) and (\ref{tempboost}), and defining 
$\del T(n^\mu) \equiv T_R(n^\mu) - \bar{T}_R$, we can write
\beq
\fr{\del T(n^\mu)}{\bar{T}_R} = \int_{\bar{t}_R}^{\bar{t}_E} \del(H_nN)dt
   + \ld.\ld(\bar{H}\del t_D + n_\mu\del t_B^{;\mu}\rd)\rd|_R^E.
\label{templin}
\eeq

   Equation (\ref{templin}) gives the observed temperature anisotropy at 
linear order in terms of the line-of-sight expansion perturbation 
$\del(H_nN)$ in an arbitrary gauge (specified by the \hsfs\ orthogonal to 
$u^\mu$) and the temporal 
displacements $\del t_B$ and $\del t_D$ required to transform from that 
arbitrary gauge to comoving and uniform density gauges.  The only 
approximations involved in \eq(\ref{templin}) are those of abrupt 
recombination and linearization.  The terms $\bar{H}\del t_D$ and 
$n_\mu\del t_B^{;\mu}$ evaluated at the LSS are sometimes called the 
``monopole'' and ``Doppler'' contributions, respectively.  The geometrical 
nature of the terms in \eq(\ref{templin}) provides a clear and unambiguous 
interpretation of the anisotropy, without reliance on coordinate-dependent 
notions such as gravitational potentials (see Ref.~\cite{hpln02} for a 
related discussion).

   Equation (\ref{templin}) is in a form that makes it easy to evaluate the 
anisotropy using any gauge for the perturbations that we choose.  First, 
for the line-of-sight integral, by linearizing the exact expression 
\eq(\ref{Hnexact}) we have
\beq
\del(H_nN) = \oo{3}\del\theta + \sig_{\mu\nu}n^\mu n^\nu + a_\mu n^\mu
           + \bar{H}\del N.
\label{delhnn}
\eeq
The geometrical quantities in this expression can be written in terms of 
the metric perturbations using \eqs(\ref{lapse}), (\ref{thetalin}), 
(\ref{sigmalin}), and (\ref{alin}), giving
\beq
\del(H_nN) = -\dot\psi + \phi_{;\mu} n^\mu + \sig_{;\mu\nu}n^\mu n^\nu
           + \half\dot H_{\mu\nu}n^\mu n^\nu.
\label{delhnnmet}
\eeq
Here $\psi$ is the curvature \pert, $\phi$ is the lapse \pert, $\sig$ is 
the shear scalar, and 
$H_{\mu\nu}$ is the tensor metric \pert.  Equation~(\ref{delhnnmet}) 
is valid in arbitrary gauges, \ie\ for arbitrary congruences 
$u^\mu$, and it is now trivial to fix the gauge, as we will see in 
Sec.~\ref{seczsgsw}.  Second, for the boundary terms in 
\eq(\ref{templin}), we can work out the required gauge 
transformations $\del t_D$ and $\del t_B$ using \eq(\ref{uedgtrns}) and 
(\ref{cgtrns}).  At the emission point $E$ those transformations are 
applied to the radiation quantities $\rho_{(\gam)}$, $P_{(\gam)}$, 
$\del\rho_{(\gam)}$, and $q_{(\gam)}$, while at the observation point 
$R$ the transformations are determined by the \hsf\ and state of motion of 
the observer chosen, as we will see.

   Since the quantity $T_R(n^\mu)$ is observable, the general expression 
\eq(\ref{templin}) must be independent of the gauge or congruence chosen, 
{\em if the point $R$ and the four-velocity of the observer 
are held constant.}  This is demonstrated explicitly in 
Appendix~\ref{ginvarapp}.  However, if the observation point and 
four-velocity are allowed to transform with the gauge transformation, then 
the anisotropies will depend on the gauge, as shown in 
Appendix~\ref{gdepapp}.  Expanding the anisotropy in terms of the spherical 
harmonics, $Y_{\ell m}(n^\mu)$, the multipole amplitudes are
\beq
a_{\ell m} \equiv \int\fr{\del T(n^\mu)}{\bar{T}_R}Y^\ast_{\ell m}(n^\mu)d\Om.
\label{multipoleexp}
\eeq
We show explicitly in Appendix~\ref{gdepapp} that only the dipole 
anisotropy $a_{1m}$ and monopole perturbation $a_{00}$ change in the latter 
case, at linear order.

\subsubsection{Recovering previous results in zero-shear or longitudinal gauge}
\label{seczsgsw}

   We will now illustrate the usefulness of the general linear expression 
for the anisotropies, \eq(\ref{templin}), by calculating 
in a particular gauge the anisotropy due to both adiabatic scalar and tensor 
sources, recovering previous results.  We require both the line-of-sight 
integral in that expression as well as the temporal displacements 
$\del t_D$ and $\del t_B$ for the boundary terms.  If we choose the congruence 
$u^\mu$ such that the scalar-derived part of the shear $\sig_{\mu\nu}$ 
vanishes at linear order, then the integrand \eq(\ref{delhnnmet}) takes a 
particularly simple form.  The frequently used longitudinal gauge has 
this property, giving
\beq
\del(H_nN) = -\dot\psi_\sig + \phi_\sig^{;\mu} n_\mu
           + \half\dot H_{\mu\nu}n^\mu n^\nu.
\eeq
In these expressions, the subscript $\sig$ indicates a zero-shear or 
longitudinal gauge quantity, and the overdot indicates the proper time 
derivative in the direction of $u^\mu$.  Therefore, using the first order 
expression
\beq
\phi_{;\mu} n^\mu = \ld.\fr{d\phi}{dt}\rd|_{v^\mu} - \dot\phi,
\eeq
we can write the integral in \eq(\ref{templin}) as
\bea
\int_{\bar{t}_R}^{\bar{t}_E} \del(H_nN)dt
   &=& \int_{\bar{t}_E}^{\bar{t}_R}
 \ld(\dot\psi_\sig + \dot\phi_\sig - \half\dot H_{\mu\nu}n^\mu n^\nu\rd)dt\nn\\
   && +\,\phi_\sig\Big|^E_R.
\label{swint}
\eea

   It is simple to calculate the displacements $\del t_D$ and $\del t_B$ 
for the case of large-scale adiabatic modes, for which the uniform 
radiation energy density and uniform total energy density \hsfs\ coincide, 
and for which the plasma-comoving and total comoving directions coincide.  
(Additionally, these surfaces and directions coincide even on small scales 
for the artificial case of a cold dark matter free universe, in the tight 
coupling approximation.)  Although $\Sls$ is defined as a surface of 
constant $\rho_{(\gam)}$, it will be easier to calculate the position of 
uniform total density surfaces.  The temporal displacement that takes us from 
zero-shear to uniform total energy density gauge is, using \eq(\ref{uedgtrns}) 
and the linearized energy constraint equation \eq(\ref{enconstr}) for 
the total energy density perturbation $\del\rho$,
\beq
\del t_D = -\fr{\del\rho_\sig}{\dot\rho}
   = -\fr{3\Hb\ld(\dot\psi_\sig + \Hb\phi_\sig\rd) - \lapp\psi_\sig}
         {12\pi G\Hb(\rho + P)}.
\label{gtrzsud}
\eeq
This expression can be applied at point $E$ on the LSS as well as at 
the reception point $R$ if we choose to place $R$ on a surface of uniform 
energy density.  The ``boost displacement'' that transforms from zero-shear 
to comoving gauge is, using \eq(\ref{cgtrns}) and the linearized 
momentum constraint equation \eq(\ref{momconstr}) for the total momentum 
density scalar $q$,
\beq
\del t_B = -\fr{\dot\psi_\sig + \Hb\phi_\sig}{4\pi G(\rho + P)}.
\label{gtrzsco}
\eeq
Again, this can be applied at both $E$ and at $R$ if we choose the observer 
to be comoving.

   Equations (\ref{swint}) to (\ref{gtrzsco}) with \eq(\ref{templin}) 
completely specify the anisotropies in terms of quantities in zero-shear or 
longitudinal gauge.  In practice it is common to make approximations.  
If we assume that the anisotropic stress is negligible (as is the case for 
matter or $\Lambda$ domination), then we have $\psi_\sig = \phi_\sig$ 
(see, \eg, \cite{mfb92}).  As explained above, the displacements $\del t_D$ 
and $\del t_B$ at the observation point $R$ only affect the observed 
monopole and dipole.  Dropping these terms, and using the background energy 
constraint, the anisotropy for $\ell > 1$ then becomes
\bea
\fr{\del T(n^\mu)}{\bar{T}_R}
   &=& \int_{\bar{t}_E}^{\bar{t}_R}
     \ld(2\dot\psi_\sig - \half\dot H_{\mu\nu}n^\mu n^\nu\rd)dt
    + \oo{3}\psi_\sig\nn\\
   &-& \fr{2}{9}\ld(\fr{3}{\Hb}\dot\psi_\sig
       + \fr{\lap}{a^2\Hb^2}\psi_\sig\rd)\nn\\
   &-& \fr{2}{3}\oo{\Hb^2}(\dot\psi_\sig + \Hb\psi_\sig)_{;\mu}n^\mu.
\label{zsgsw}
\eea
All quantities outside the integral are to be evaluated at the point 
on the LSS corresponding to viewing direction $-n^\mu$.  This expression 
agrees precisely with \eq(4.7) in Ref.~\cite{cl98}, for the case of adiabatic 
perturbations, including even a term the authors of \cite{cl98} 
describe as arising from ``subtle gauge effects.''

   Making further simplifications, we have $\dot\psi_\sig = 0$ at 
last scattering if the pressure vanishes exactly there.  
If we consider 
anisotropies on the largest angular scales, sourced by modes with comoving 
wavenumber $k \ll a\bar{H}$, we can drop all the gradient terms in 
\eq(\ref{zsgsw}).  The result is
\beq
\fr{\del T(n^\mu)}{\bar{T}_R}
   = \int_{\bar{t}_E}^{\bar{t}_R}
     \ld(2\dot\psi_\sig - \half\dot H_{\mu\nu}n^\mu n^\nu\rd)dt
   + \oo{3}\psi_\sig(E),
\label{zsgswls}
\eeq
in agreement with the well-known result for the Sachs-Wolfe effect due to 
large-scale scalar and tensor sources.  Note that this result includes a 
part that is evaluated at the boundary $E$, which comes from both the 
temporal displacement $\del t_D(E)$ and the integral in \eq(\ref{templin}).  
The remainder of that integral, which cannot be placed at the boundary, 
appears in \eq(\ref{zsgswls}) as a contribution that is due to physical 
metric fluctuations along the line of sight.  The scalar part of this 
contribution is known as the integrated Sachs-Wolfe effect.  During matter 
domination $\dot\psi_\sig = 0$, so all scalar effects of the perturbed 
expansion along the line of sight can be placed at the boundary.

\section{Transfer functions}
\label{secmonodi}

\subsection{Dipole}
\label{dipsubsec}

   Next we will use the results derived so far to perform a careful 
calculation of the CMB dipole due to scalar sources.  More precisely we 
will calculate the dipole power, or the variance in the dipole anisotropy 
$a_{1m}$, namely
\beq
C_1 \equiv \bra|a_{1m}|^2\ket - |\bra a_{1m}\ket|^2,
\label{C1def}
\eeq
over realizations of the assumed Gaussian random primordial fluctuations.  
(The independence of $C_1$ on $m$ will follow from statistical isotropy.)  
For the dipole the choice of frame for the observer at $R$ is critical, 
since a boost at the observation point changes the dipole according to 
\eq(\ref{dipoleboost}).  For example, if we choose the observer's frame 
to be comoving with radiation, so that the radiation momentum density 
$q^\mu_{(\gam)}$ vanishes, then trivially the observed dipole vanishes.  
Indeed, combining \eqs(\ref{qgtrns}) and (\ref{dipoleboost}) we have
\beq
\oo{3}\sum_m|a_{1m}|^2 =\fr{\pi}{4\rho_{(\gam)}^2}q^\mu_{(\gam)}q^{(\gam)}_\mu,
\label{dipoleflux}
\eeq
so that the magnitude of the observed dipole in any frame is proportional 
to the radiation flux observed in that frame.  We will adopt the most 
natural and physically best-motivated choice, namely the frame comoving with 
matter (essentially the total comoving frame at late times) for the 
calculation of $C_1$.  With this choice of frame, \eq(\ref{dipoleflux}) 
has the simple interpretation that the observed dipole is a measure 
of how well worldlines comoving with radiation and matter coincide.  

   We begin with the general linear result, \eq(\ref{templin}).  All of the 
parts of this equation were carefully calculated in zero-shear gauge for a 
comoving observer in Sec.~\ref{seczsgsw}.  We can ignore the monopole 
contribution $\del t_D(R)$ since it only affects the $\ell = 0$ mode.  
The displacements $\del t_D(E)$ and $\del t_B$ were calculated in 
\eqs(\ref{gtrzsud}) and (\ref{gtrzsco}).  Combining these results with 
\eq(\ref{swint}) for the line-of-sight integral, and ignoring anisotropic 
stress at all times (so $\psi_\sig = \phi_\sig$), assuming matter domination 
at last scattering (so $\dot\psi_\sig(E) = 0$), and finally ignoring the 
term $\lapp\psi_\sig(E)$, we have\begin{widetext}
\beq
\fr{\del T(n^\mu)}{\bar{T}_R}
   = \oo{3}\psi_\sig(E) - \fr{2}{3}\fr{n_\mu\psi_\sig^{;\mu}(E)}{\Hb_E}
   + \ld(\fr{5}{3}g_R^{-1} - 1\rd)\fr{n_\mu\psi_\sig^{;\mu}(R)}{\Hb_R}
   + 2\int_E^R \dot\psi_\sig dt.
\label{dideltt}
\eeq
Here we have used \eq(\ref{gdefn}) for the growth function $g_R \equiv g(t_R)$ 
and the relation \eq(\ref{greln}) to simplify the expression.  The 
approximation of matter domination at last scattering (which implies zero 
anisotropic stress) results in errors in $C_\ell$ for small $\ell$ on the 
order of $10\%$ \cite{hs95}.  Neglecting the term $\lapp\psi_\sig(E)$ is 
entirely justified considering that \eq(\ref{gtrzsud}) for $\del t_D(E)$ 
employed the approximation that the uniform radiation energy density and 
uniform total energy density \hsfs\ coincide, which is only valid on 
large scales (scales that were super-Hubble at last scattering).  The 
contribution to the dipole from last scattering will be dominated by these 
large scales, as we will see.

   To calculate the variance of the dipole, it will be helpful to expand 
the function $\psi_\sig(E)$ in spherical harmonics as
\beq
\psi_\sig(E) = -\fr{3}{5}\sqrt{\fr{2}{\pi}} \int dk\,k T(k) \sum_{\ell m} 
                \Rpr_{\ell m}(k) \jl(k\rls)Y_{\ell m}(-n^\mu).
\label{phiexpe}
\eeq
Here $\jl$ is the spherical Bessel function of the first kind, $\rls$ is 
the comoving radius of the LSS, $k$ is the comoving wave number, and $T(k)$ 
is the transfer function defined in \eq(\ref{rtransfnc}).  With the aim of 
expressing $C_1$ in terms of the primordial spectrum $\PR(k)$, we have used 
\eq(\ref{psiR}) to write $\psi_\sig$ in terms of the primordial comoving 
curvature perturbation $\Rpr$.  Similarly we will need the expansion of 
the ``Doppler'' contributions at $E$ and $R$,
\beq
n_\mu\psi^{;\mu}(E) = \fr{3}{5}\sqrt{\fr{2}{\pi}}\oo{a_E} \int dk\,k^2 T(k)
     \sum_{\ell m} \Rpr_{\ell m}(k) \jl'(k\rls)Y_{\ell m}(-n^\mu),
\label{dipexpe}
\eeq
and
\bea
n_\mu\psi^{;\mu}(R)
   &=& \lim_{r \ra 0}\fr{3}{5}\sqrt{\fr{2}{\pi}}\fr{g_R}{a_R}\int dk\,k^2 T(k)
       \sum_{\ell m} \Rpr_{\ell m}(k) \jl'(kr)Y_{\ell m}(-n^\mu)\\
   &=& \oo{5}\sqrt{\fr{2}{\pi}}\fr{g_R}{a_R}
       \int dk\,k^2 T(k) \sum_m \Rpr_{1m}(k) Y_{1m}(-n^\mu),
\label{dipexpr}
\eea
where the prime indicates differentiation with respect to the argument, 
and we have used the relation $j'_\ell(0) = \del^1_\ell/3$.  
Inserting these expressions into \eq(\ref{dideltt}), we can evaluate the 
dipole anisotropy using \eq(\ref{multipoleexp}) (with $\ell = 1$) and the 
orthonormality of the $Y_{\ell m}$.  This gives
\beq
a_{1m} = -\oo{5}\sqrt{\fr{2}{\pi}} \int dk\,k \,\Rpr_{1m}(k) T(k) T_1(k),
\eeq
where $T_1(k)$ is a new transfer function, called the {\em dipole transfer 
function,} and defined by
\beq
T_1(k) \equiv j_1(k\rls) + \fr{2k}{a_E\Hb_E}j'_1(k\rls)
   + \ld(g_R - \fr{5}{3}\rd)\fr{k}{a_R\Hb_R} + 6\int_E^R\dot g(t)j_1[kr(t)]dt.
\label{ditransfnc}
\eeq
\end{widetext}Here $r(t)$ is the comoving radial coordinate of the light 
ray as a function of the time coordinate.  Finally, using \eq(\ref{C1def}) 
and the statistical relation \eq(\ref{Rstat}) 
for the primordial power spectrum $\PR(k)$, we find for the dipole power
\beq
C_1 = \fr{4\pi}{25}\int\fr{dk}{k} \PR(k)T^2(k)T_1^2(k).
\label{dipower}
\eeq

   Note the third term in \eq(\ref{ditransfnc}), which is proportional to 
$k$ and dominates on small scales today.  It might appear that this term 
would lead to an ultraviolet divergence for the dipole, but 
\eq(\ref{dipower}) is rendered finite by the function $T(k)$, which decays 
like $k^{-2}$ on small scales.  This results in a dipole amplitude 
$\sqrt{C_1} \sim 10^{-3}$ for standard cosmological parameters.

\subsection{Monopole}
\label{secmono}

   The monopole temperature perturbation is the $\ell = 0$ component of 
\eq(\ref{multipoleexp}), \ie
\beq
a_{00} = \oo{\sqrt{4\pi}}\int\fr{\del T(n^\mu)}{\bar{T}_R} d\Om,
\label{monodef}
\eeq
and its variance over realizations of the primordial fluctuations will be 
called $C_0$.  As discussed above \eq(\ref{bgndtfreedom}), the monopole so 
defined is ambiguous in that different choices of the background times 
$\bar t_E$ and $\bar t_R$ lead to different $a_{00}$.  This well-known 
freedom is equivalent to our inability to uniquely separate a ``background'' 
CMB temperature from the observed mean temperature, which would be required 
to define a monopole perturbation.  Hence the monopole perturbation in 
\eq(\ref{monodef}) cannot be observable.

   Nevertheless, it is still possible to sensibly define a monopole and 
consider its theoretical properties.  Recall that for the dipole it was 
necessary to fix the observer's frame by a physical prescription (namely 
that it be comoving with matter).  Analogously, to define a monopole the 
key point is that we must fix, by a physical prescription, the spacelike 
\hsf\ on which we place the observer.  Clearly there is freedom in how we 
choose this slice.  Recall 
that an observer chosen to comove with radiation observes no dipole.  The 
analogous situation with the monopole is an observer placed on a 
uniform radiation energy density \hsf, for which the calculated $C_0$ 
must vanish.  Analogously to \eq(\ref{dipoleflux}) for the dipole, we 
can write
\beq
|a_{00}|^2 =\fr{\pi}{4\rho_{(\gam)}^2}(\del\rho_{(\gam)})^2,
\label{monoflux}
\eeq
where $\del\rho_{(\gam)}$ is the radiation energy density perturbation 
on the same slicing used to define the monopole $a_{00}$.  Therefore the 
monopole defined with respect to any slicing is proportional to the 
radiation density perturbation for the same slicing.

   The simplest 
and most natural choice of slice on which to place the observer is that 
of uniform matter (essentially total) energy density.  This choice is 
simplest because it requires knowledge of just the local density, $\rho$, 
which acts as a clock.  It is natural because, as we will see, it exhibits 
close analogy with the dipole.  Comoving slices could also be used, 
although their construction as \hsfs\ orthogonal to the comoving worldlines 
is more elaborate \cite{notecomove}.  
Thus, while the dipole defined above is a measure of the degree to which 
comoving radiation and matter worldlines coincide, the monopole defined 
here has the simple interpretation as a measure of how well \hsfs\ of 
uniform radiation and matter energy density coincide \cite{noteergodic}.  
The arbitrariness 
in $a_{00}$ mentioned above (due to the freedom to choose the background 
times) only amounts to a constant shift to $a_{00}$, so its variance is 
unchanged.  This is why, although the monopole perturbation $a_{00}$ is 
ambiguous and unobservable for any single observer, the {\em variance} 
$C_0$ is still well defined theoretically (and could, in principle, be 
approximated through observations \cite{notemonomeas}).

   It should be clear immediately that there is a serious problem with 
attempting to define the variance of the CMB temperature on a uniform 
matter density slice.  Namely, matter has of course entered the nonlinear 
stage on small scales, and hence \hsfs\ of constant matter density cannot 
actually be defined.  Nevertheless, smoothing over small scales can recover 
meaningful linear results, at the expense of further ambiguity in the 
form of the smoothing scale.  This issue arises because in calculating 
the monopole as defined here, we will see that the dominant contribution 
will be given by the variance $\bra(\del\rho/\rho)^2\ket$ evaluated at $R$, 
which is dominated by small scales (where the slicing is irrelevant).  The 
important practical aspect of this brief examination of the monopole 
will actually be in understanding the effects of long-wavelength sources, 
for which the difficulties with small scales do not arise.

   Proceeding as with the dipole above, the relevant contributions to 
the general linear result \eq(\ref{templin}) give\begin{widetext}
\beq
\fr{\del T(n^\mu)}{\bar{T}_R}
   = \oo{3}\psi_\sig(E) - \fr{2}{3}\fr{n_\mu\psi_\sig^{;\mu}(E)}{\Hb_E}
   + \ld(\fr{5}{3}g_R^{-1} - 2\rd)\psi_\sig(R)
   - \fr{2}{9\Om_m}\fr{\lap}{a_R^2\Hb_R^2}\psi_\sig(R)
   + 2\int_E^R \dot\psi_\sig dt.
\label{monodeltt}
\eeq
We can ignore the ``Doppler'' contribution $\del t_B$ at $R$ since it 
only affects the $\ell = 1$ mode, and the same approximations have been 
applied here as for the dipole \eq(\ref{dideltt}).

   To calculate the variance of the monopole 
we will need, in addition to \eqs(\ref{phiexpe}) and (\ref{dipexpe}), 
the ``monopole'' term evaluated at the reception point $R$,
\bea
\psi_\sig(R) &=& -\fr{3}{5}\sqrt{\fr{2}{\pi}}g_R \int dk\,k T(k)
                 \sum_{\ell m} \Rpr_{\ell m}(k) \jl(0)Y_{\ell m}(n^\mu)\\
       &=& -\fr{3}{5}\sqrt{\fr{2}{\pi}}g_R \int dk\,k T(k)\Rpr_{00}(k) Y_{00},
\eea
using the relation $\jl(0) = \del^0_\ell$.  Proceeding as for the dipole 
case, we find
\beq
a_{00} = -\oo{5}\sqrt{\fr{2}{\pi}} \int dk\,k \Rpr_{00}(k)T(k)T_0(k),
\eeq
where $T_0(k)$ is the monopole transfer function defined by
\beq
T_0(k) \equiv j_0(k\rls) + \fr{2k}{a_E\Hb_E}j'_0(k\rls) + 5 - 6g_R
          + \fr{2}{3}\fr{g_R}{\Om_m}\fr{k^2}{a_R^2\Hb_R^2}
          + 6\int_E^R \dot g(t)j_0[kr(t)]dt.
\label{monotransfnc}
\eeq
\end{widetext}Finally, using the statistical relation \eq(\ref{Rstat}) to 
evaluate the variance we find
\beq
C_0 = \fr{4\pi}{25}\int\fr{dk}{k} \PR(k)T^2(k)T_0^2(k).
\label{monomatter}
\eeq
Note the presence of the Laplacian term ($\propto k^2$) in $T_0(k)$, which 
will dominate 
on small scales, and implies that by calculating $C_0$ we are essentially 
calculating the (effectively gauge independent) matter variance 
$\bra(\del\rho/\rho)^2\ket$ at the observation point.  In this monopole 
case the resulting divergence is too strong to be saved by the transfer 
function $T(k)$.  Therefore we expect the variance $C_0$ to 
diverge on small scales, which is simply a reflection of the nonlinear 
nature of matter fluctuations on small scales today, as predicted above.  
That is, the monopole variance as we have defined it cannot be quantified.  
Again, the importance of \eq(\ref{monotransfnc}) will lie in its 
long-wavelength behaviour.

\section{Long-wavelength behaviour}
\label{seclongwl}

\begin{figure}[ht]\begin{center}
\includegraphics[width=\columnwidth]{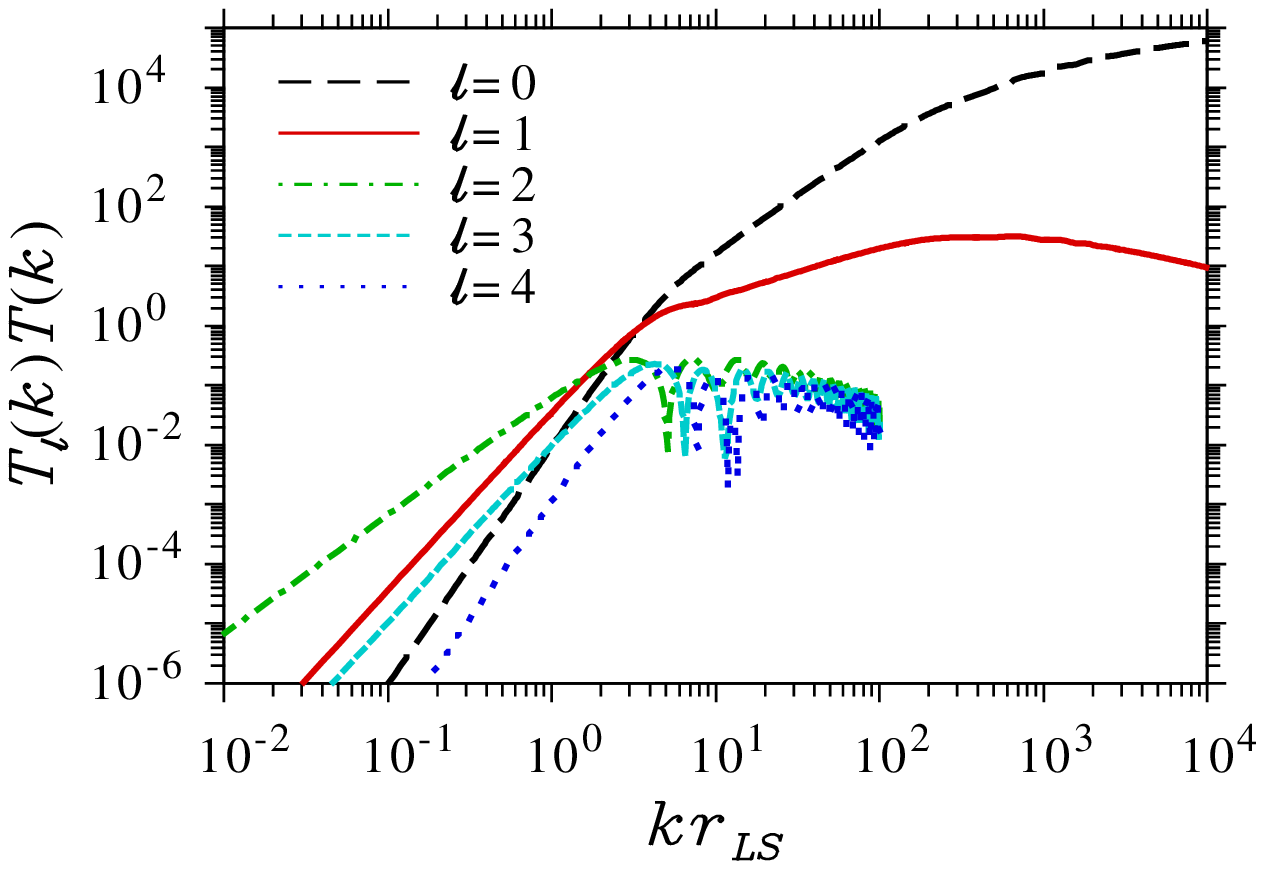}
\caption{Dipole transfer function $T_1(k)T(k)$ for a comoving observer 
(solid curve).  For comparison, the monopole transfer functions for an 
observer on a uniform energy density slice, $T_0(k)T(k)$, and the transfer 
functions for $\ell = 2$, $3$, and $4$ are also shown.  Absolute values are 
plotted.  The scales $k\rls = 10^2$ and $k\rls = 1$ correspond roughly to 
the Hubble scales at last scattering and today, respectively.}
\label{transfnc}
\end{center}\end{figure}

   Recall that all of our approximations have been good on very large scales.  
In this section we examine the long-wavelength limit of the 
$T_1(k)$ and $T_0(k)$ transfer functions.  The dipole transfer function 
for a comoving observer, 
\eq(\ref{ditransfnc}), is plotted in \fig\ref{transfnc}, together with the 
monopole function for an observer on a uniform energy density slice, 
\eq(\ref{monotransfnc}), and the transfer functions for $\ell = 2$, $3$, 
and $4$.  The transfer functions for $\ell > 1$ can be calculated from 
\eq(\ref{dideltt}) in exactly the same way as for the dipole, with the result
\beq
T_\ell(k) = j_\ell(k\rls) + \fr{2k}{a_E\Hb_E}j'_\ell(k\rls)
   + 6\int_E^R\dot g(t)j_\ell[kr(t)]dt.
\label{hitransfnc}
\eeq
Note that the large-scale approximations involved in \eq(\ref{dideltt}) 
imply that this expression is only valid for scales that are super-Hubble 
at last scattering.  [The transfer functions $T_1(k)$ and $T_0(k)$ {\em are} 
valid for small scales, since for large $k\rls$ the second-to-last terms in 
both \eqs(\ref{ditransfnc}) and (\ref{monotransfnc}) dominate.  These 
terms are generated {\em locally} at the observation point $R$.]  A value 
of $\Omega_\Lam = 0.77$ today was used for all of these calculations.

   Using asymptotic forms for the Bessel functions \cite{as72}, we can show 
from \eq(\ref{hitransfnc}) that $T_\ell(k)$ should decay like $k^\ell$ as 
$k\rls \to 0$, for $\ell > 1$.  This is verified in \fig\ref{transfnc}.  
However, the figure also shows that $T_1(k)$ does not decay like $k$ for 
small $k$; instead it decays like $k^3$, which is {\em faster} than the 
decay rate of $T_2(k)$.

   To examine the behaviour of the transfer function $T_1(k)$ in the limit 
$k \ra 0$, we can use the small-argument approximations to the Bessel 
functions \cite{as72} to give
\bea
T_1(k) &=& \ld[\ld(g_R - \fr{5}{3}\rd)\oo{a_R\Hb_R} - \fr{5}{3}\eta_R
            + 2\int_0^{\eta_R} g(\eta) d\eta\rd]k\nn\\
       && +\,\O(k\rls)^3,
\label{T1lam}
\eea
where the conformal time $\eta$, defined via $d\eta = dt/a$, has been 
used to simplify the expression.  In writing \eq(\ref{T1lam}), we have 
used the approximation $g(t_E) = 1$, although for our numerical 
calculations we have evaluated \eq(\ref{ditransfnc}) without 
approximation.  It is not obvious from \eq(\ref{T1lam}) 
that the $\O(k)$ term, in square brackets, vanishes.  However, the 
numerical calculation of \fig\ref{transfnc} shows that the $\O(k)$ term 
does indeed vanish, as pointed out above.

\begin{figure}[ht]\begin{center}
\includegraphics[width=\columnwidth]{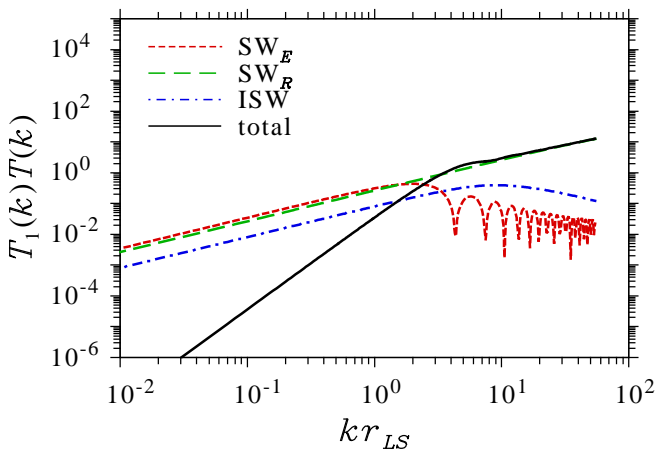}
\caption{Dipole transfer function $T_1(k)T(k)$ for a comoving observer.  
Absolute values of individual contributions from Sachs-Wolfe terms 
evaluated at emission, $E$, and at the observation point, $R$, as well as 
the line-of-sight ISW contribution, are indicated.}
\label{ditrans}
\end{center}\end{figure}

   This is illustrated in greater detail in \fig\ref{ditrans}.  There we 
have plotted separately the first two terms in \eq(\ref{ditransfnc}), which 
originate at the LSS (called ${\rm SW}_E$ in the plot), the third term 
in \eq(\ref{ditransfnc}), which comes from the observation point $R$ 
(${\rm SW}_R$), and the line-of-sight ISW term.  
We can see that each component separately 
does scale like $k$ on large scales, in particular the LSS term 
${\rm SW}_E$ scales like the predicted $k^\ell$, but the sum demonstrates 
that the $\O(k)$ terms all cancel.  It is the local contribution, 
${\rm SW}_R$, not present for $\ell > 1$, which enables the 
cancellation.  (Of course the individual SW and ISW components are not 
separately observable!)  It is possible to show analytically this $\O(k^3)$ 
dependence 
in the special case of a cosmological-constant free Einstein-de~Sitter 
universe.  Then \eq(\ref{T1lam}) becomes
\beq
T_1(k) = -\fr{k^3}{30}\ld(\rls^3 + 3\rls^2\etals\rd) + O(k^5).
\eeq

   Figure~\ref{ditrans} also illustrates that the source of the small-scale 
increase in $T_1(k)$ is due to the local contribution at $R$, as explained 
above.  The dipole transfer function, \eq(\ref{ditransfnc}), suggests a 
divergence in $C_1$ like a power of $k$ on small scales, although this is 
moderated by the transfer function $T(k)$.  The result is that the observed 
dipole amplitude is a factor $\sim 10^2$ larger than the other multipoles.  
Figure~\ref{transfnc} actually illustrates this directly:  \eq(\ref{dipower}) 
generalizes to
\beq
C_\ell = \fr{4\pi}{25}\int\fr{dk}{k} \PR(k)T^2(k)T_\ell^2(k)
\eeq
for all $\ell \ge 0$.  Therefore the expected multipole amplitude is 
simply proportional to the area under the appropriate curve in 
\fig\ref{transfnc} [assuming a 
nearly scale-invariant primordial spectrum $\PR(k)$].  Geometrically, the 
comoving matter and radiation worldlines coincide on large scales, and begin 
to diverge strongly on small scales.  The contribution to the dipole from 
last scattering, ${\rm SW}_E$, will vary smoothly if the observer's position 
is varied, whereas the local contribution, ${\rm SW}_R$, will vary greatly.

   These results indicate that the dipole defined with respect to the 
comoving frame receives strongly suppressed contributions from super-Hubble 
modes.  This was in fact noted some time ago \cite{turner91} in relation 
to the Grishchuk-Zel'dovich effect \cite{gz78}, but in the context of a 
matter-dominated universe.  In fact, it was claimed in \cite{turner91} that 
this cancellation would not persist in the presence of a cosmological 
constant, so that super-Hubble modes would have an observable imprint on 
the CMB, a claim which we have now demonstrated to be incorrect.

\begin{figure}[ht]\begin{center}
\includegraphics[width=\columnwidth]{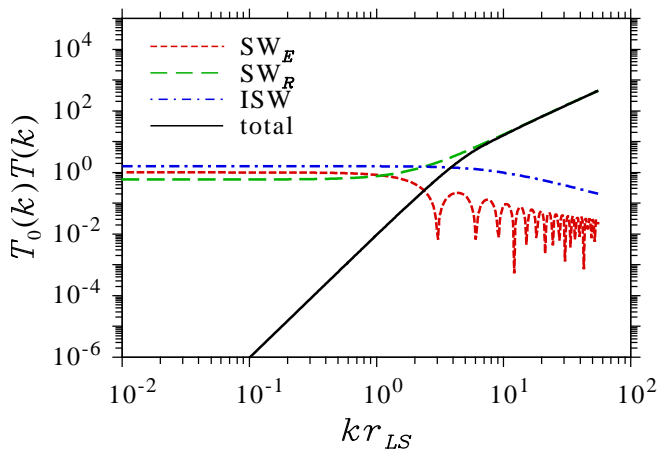}
\caption{Monopole transfer function $T_0(k)T(k)$, for an observer on a 
uniform energy density slice.  Absolute values of individual contributions 
from Sachs-Wolfe terms evaluated at emission, $E$, and at the observation 
point, $R$, as well as the line-of-sight ISW contribution, are indicated.}
\label{monotrans}
\end{center}\end{figure}

   As we did for the dipole, we can examine the behaviour of the 
monopole transfer function $T_0(k)$ in the limit $k \ra 0$, giving
\bea
T_0(k) &=& \ld[-\oo{6}\rls^2 - \fr{2\rls}{3a_E\Hb_E} - \fr{2}{3a_E^2\Hb_E^2}
   + \fr{2g_R}{3\Omega_{m,R}a_R^2\Hb_R^2}\rd.\nn\\
   && -\,\ld.\int_E^R\dot g(t)r^2(t)dt\rd]k^2 + \O(k^4),
\label{T0lam}
\eea
where $\Omega_m$ is the standard matter density parameter.  In this 
monopole case, we have thus shown analytically that the leading-order 
terms in the transfer function, \eq(\ref{monotransfnc}), which go like 
$k^0$, do cancel in the presence of a cosmological constant.  However, 
\fig\ref{transfnc} shows that an even 
stronger result holds in the monopole case:  the $\O(k^2)$ terms 
in \eq(\ref{T0lam}) actually cancel as well!  Again, this is shown in 
greater detail in \fig\ref{monotrans}, where 
it is apparent that an exquisite cancellation occurs in the individual 
components, which scale like $k^0$ on large scales, to give the $\O(k^4)$ 
total transfer function on large scales.  Figure~\ref{monotrans} also 
illustrates the 
local source, ${\rm SW}_R$, of the strong small-scale divergence 
discussed above.  Geometrically, the \hsfs\ of uniform matter and 
radiation density coincide on large scales, and begin to very strongly 
diverge on small scales.  This divergence makes it impossible to quantify 
the total power $C_0$ with our definition of the monopole.

\section{Discussion}
\label{discusssec}

   Our result in Sec.~\ref{seclongwl} that the dipole and monopole 
receive suppressed contributions from large scales, even in the presence 
of a dominant cosmological constant, strongly suggests that the 
cancellations involved are not accidental.  To understand the 
origin of this behaviour, consider first the monopole case.  Physically, 
the suppression as $k \ra 0$ in the monopole transfer function shown in 
Figs.~\ref{transfnc} and \ref{monotrans} means that surfaces of uniform 
total and radiation energy density coincide on the largest scales, 
according to our definition of the monopole in Sec.~\ref{secmono}.  But 
this is just the statement of adiabaticity: an adiabatic matter-radiation 
fluid is characterized by the condition
\beq
\fr{\del\rho}{\dot\rho} = \fr{\del\rho_{(\gam)}}{\dot\rho_{(\gam)}},
\label{adcond}
\eeq
on the largest scales.  Therefore, \eq(\ref{uedgtrns}) shows that the 
same gauge transformation takes us to both constant total matter and 
constant radiation 
\hsfs; in other words, those surfaces must coincide.  It is important 
to point out that adiabatic long-wavelength modes remain adiabatic 
under evolution \cite{mfb92}, and indeed the comoving curvature 
\pert\ $\R$ remains constant in time (see, \eg, \cite{ll00}).  This 
trivial evolution on super-Hubble scales means that when constant 
matter and radiation density surfaces coincide on large scales at last 
scattering, due to adiabatic initial conditions, they must also coincide today.

   An analogous situation holds for the dipole.  In this case, 
the adiabaticity condition implies that, on large scales, the 
radiation and total matter comoving worldlines coincide (\ie, there 
is no ``peculiar velocity'' isocurvature mode between the two components).  
According to our definition of the dipole in Sec.~\ref{dipsubsec}, 
this is simply the statement that the dipole is suppressed on large 
scales.

   This leads us to the important conclusion that this insensitivity 
to long-wavelength sources must apply regardless of the matter content 
of the Universe, as long as adiabaticity holds.  Therefore the suppression 
we found for the specific case of cosmological constant (or 
Einstein-de~Sitter) universes must in fact occur in general.  Note 
that this may have relevance to very recent discussions regarding 
a potential power asymmetry in the CMB \cite{ekc08}.  The exquisite 
cancellations visible at large scales in Figs.~\ref{ditrans} and 
\ref{monotrans} between the SW and ISW components illustrate a 
previously unrecognized relation between the two, which is enforced by 
the condition of adiabaticity.

   In brief, the dipole and monopole as defined here are just measures 
of departures 
from the adiabaticity condition, \eq(\ref{adcond}), which generally occur on 
small scales.  Are these results sensitive to the definitions of dipole 
and monopole used?  As long as the dipole and monopole are defined 
{\em physically}, \ie\ in relation to locally measureable quantities, 
then the results must still hold.  An important example which does {\em not} 
satisfy this criterion is the zero-shear, or longitudinal gauge, since it 
is not defined in terms of local, observable matter quantities.  If the 
dipole (or monopole) is defined with respect to a zero-shear frame, then 
it may appear from theoretical calculations that the dipole {\em is} 
sensitive to long-wavelength modes.  However, this sensitivity cannot be 
observable, 
since zero-shear frames cannot be {\em uniquely} constructed locally.  
(A linear boost or ``tilt'' of a zero-shear frame is still a zero-shear 
frame.)  Thus great care must be taken when considering behaviour on large 
scales with longitudinal gauge.

   Indeed, another way to understand this result is to realize that, 
in the limit $k/(aH) \ra 0$, an adiabatic \pert\ mode locally becomes 
essentially pure gauge, 
and can be removed within any {\em sub-Hubble} region by a simple 
boost coordinate transformation.  Such a mode is indistinguishable 
locally from a homogeneous background, and hence cannot have any 
observational consequences such as a dipole anisotropy \cite{unruh98}.

   Finally, we note that when the condition of adiabaticity is relaxed, 
then our conclusions no longer hold \cite{langlois96}.  In the presence of 
isocurvature 
\perts, it is possible that a {\em physically} defined dipole be sensitive 
to super-Hubble modes, since the extra freedom allows for 
a relative tilt between comoving matter and radiation comoving worldlines 
on large scales.

   While we have emphasized here the consequences of adiabaticity, it is 
hoped that other applications will follow from the exact formalism 
for CMB anisotropies that we have developed.  One possibility is the 
evaluation of 
anisotropies, in particular the ISW effect, in void models of acceleration 
\cite{celerier99} (see \cite{enqvist08} for a brief review), which have not 
yet been confronted with observations at the perturbative level 
\cite{zibin08}.  Note however that the present approach is not limited to 
calculating CMB anisotropies, and that more generally it is applicable 
to calculating redshifts in arbitrary spacetimes.  Potential uses include 
calculating the redshift-luminosity distance relation in perturbed 
spacetimes (see, \eg, \cite{hg06}).

   {\em Note added:}  When this work was essentially complete, a related 
paper appeared \cite{eck08}, which appears to support our conclusion that 
long-wavelength \perts\ cannot effect the CMB dipole.

\begin{acknowledgments}
This research was supported by the Natural Sciences and Engineering Research 
Council of Canada.  We thank Tom Waterhouse for many useful discussions.
\end{acknowledgments}

\appendix
\section{Metric-based approach}
\label{secmetappr}

   In this appendix, we will collect together several elements of the 
metric-based approach to cosmological \perts\ that will be useful in 
describing the CMB anisotropies at linear order (see, \eg, \cite{mfb92} for a 
review of metric-based \pert\ theory).   In the metric approach a set of 
coordinates are defined in the spacetime by a foliation into spacelike \hsfs\ 
of constant $t$, $\Sig_t$, and a threading into timelike worldlines of 
constant spatial coordinates $x^i$, where latin indices run from $1$ to 
$3$.  The gauge freedom of \pert\ theory is related to our freedom to 
choose such a slicing and threading.  
However, it can be shown (see \cite{bardeen88,hwang91}) that at linear 
order, {\em physical} \perts\ in FRW backgrounds are gauge invariant under 
changes in the threading.  Therefore the gravitational dynamics can be 
entirely expressed in terms of spatially gauge-invariant quantities.  The 
reason for this invariance is simply the homogeneity of the background 
spacetime, and its importance is that it means that there is effectively 
just a single degree of gauge freedom on FRW backgrounds, namely the 
temporal position of $\Sig_t$ at each event.  Thus, to simplify expressions 
to follow, we will choose the congruence of coordinate threads (with 
tangents $u^\mu$) to be orthogonal to the $\Sig_t$, so that the shift 
vector (metric component $g_{0i}$) vanishes.  These spatial coordinates 
are comoving at zeroth order, but may depart from comoving at first order.  
With this choice, there is a direct correspondence between the gauge 
freedom (\ie\ the freedom to choose the slicing) in the metric formalism, 
and the freedom to choose the congruence in the covariant formalism.  
Evaluating the metric using \eq(\ref{spatmetdef}) in the chosen coordinates 
we have
\beq
g_{00} = -N^2, \quad g_{ij} = h_{ij},
\eeq
where $N$ is the lapse function for the slicing.  We consider scalar and 
tensor \perts\ only, as vectors are ordinarily thought to be cosmologically 
irrelevant.

   The spacetime is completely described by the lapse function, the 
intrinsic curvature of the spatial metric $h_{\mu\nu}$, and the extrinsic 
curvature of the $\Sig_t$.  We define the lapse perturbation $\phi$ through
\beq
N \equiv 1 + \phi.
\label{lapse}
\eeq
The only part of the intrinsic curvature of $h_{\mu\nu}$ that we will need 
is the perturbed Ricci scalar, $\del\Rs$, for the spatial slices $\Sig_t$, 
which defines the curvature perturbation $\psi$ via the relation
\beq
\del\Rs \equiv \fr{4}{a^2}\lap\psi.
\eeq
Here $\lap/a^2 \equiv D_\mu D^\mu$, for background scale factor $a$, so 
$\lap$ is the comoving Laplacian.  The extrinsic curvature of the slicing 
is specified by the expansion and shear of the normal congruence to the 
$\Sig_t$, and is related to the spatial metric through \cite{wald84}
\beq
\oo{3}\theta h_{ij} + \sig_{ij} = \half\dot h_{ij},
\label{extrcurv}
\eeq
where the overdot represents the proper time derivative along $u^\mu$.  
At linear order the trace of this equation gives for the expansion 
perturbation
\beq
\del\theta = -3\bar{H}\phi - 3\dot\psi + \lapp\sig,
\label{thetalin}
\eeq
where $\bar{H}$ is the background Hubble rate.  The shear scalar $\sig$ 
describes the scalar-derived part of the shear, with
\beq
\sig_{ij} = D_i D_j\sig - \oo{3a^2}\lap\sig h_{ij} + \half\dot H_{ij},
\label{sigmalin}
\eeq
where $H_{ij}$ is the transverse and traceless (tensor) part of $h_{ij}$, 
and we have ignored the vector-derived part of the shear.  One further 
important quantity is the acceleration of the worldlines 
normal to the $\Sig_t$, \eq(\ref{acceldef}).  It is related to the lapse 
through the exact expression
\beq
a_\mu = \oo{N}D_\mu N,
\eeq
which at linear order becomes
\beq
a_\mu = D_\mu\phi.
\label{alin}
\eeq

   The various quantities defined above can be related to the explicit 
component form of the metric at linear order (ignoring the vector part),
\beq
\del g_{00} = -2\phi, \quad 
\del g_{ij} = a^2(-2\psi\gam_{ij} + 2E_{,ij} + H_{ij}),
\eeq
where $\gam_{ij}$ is the background spatial metric, and the trace-free 
scalar part of \eq(\ref{extrcurv}) gives $\sig = a^2\dot E$.

   Once the slicing $\Sig_t$ is specified, \ie\ the time coordinate is 
chosen, then the perturbation in any exact quantity $X(x^i, t)$ is fixed via
\beq
\del X(x^i, t) = X(x^i, t) - \bar{X}(t),
\eeq
where $\bar{X}(t)$ is the homogeneous background value.  Therefore our 
freedom to vary the $\Sig_t$ (equivalently to vary the orthogonal 
congruence $u^\mu$) results in an inherent ambiguity in our ability to 
specify any perturbation.  This temporal gauge freedom can be used to simplify 
calculations by choosing an appropriate congruence, as we see with 
the Sachs-Wolfe calculation in Sec.~\ref{seczsgsw}.  It is 
straightforward to calculate the change in any perturbation under a gauge 
transformation $t \ra t - \del t$ (see, \eg, \cite{hwang91}).  A few 
results that we will need are, at linear order,
\bea
\del\rho &\ra& \del\rho + \dot\rho \del t,\label{rhogtrns}\\
       q &\ra& q - (\rho + P) \del t,\label{qgtrns}\\
    \psi &\ra& \psi - \Hb\del t,\label{psigtrns}
\eea
where $q$ is the scalar from which the momentum density $q^\mu$ is derived, 
through $q^\mu = D^\mu q$.  These results imply that the \gtrn\ required 
to go from an arbitrary initial gauge to uniform energy density gauge, 
defined by $\del\rho = 0$, is
\beq
\del t = -\fr{\del\rho}{\dot\rho},
\label{uedgtrns}
\eeq
and the transformation that takes an arbitrary initial gauge into comoving 
gauge, defined by $q = 0$, is
\beq
\del t = \fr{q}{\rho + P}.
\label{cgtrns}
\eeq
Unless otherwise stated, all expressions in this work will be presented 
in an unspecified gauge, \ie\ they will apply to arbitrary gauges.

   We will also need the Einstein constraint equations in order to 
relate the matter perturbations to the metric perturbations.  Projecting 
the Einstein equations twice along $u^\mu$ and linearizing, we find the 
energy constraint,
\beq
3\Hb(\dot\psi + \Hb\phi) - \lapp(\psi + \Hb\sig) = -4\pi G\del\rho.
\label{enconstr}
\eeq
Similarly, projecting once with $u^\mu$ and once with $h\ud{\mu}{\nu}$ gives 
the linearized (scalar) momentum constraint
\beq
\dot\psi + \Hb\phi = -4\pi Gq.
\label{momconstr}
\eeq

   It is conventional to express the primordial power spectrum in terms 
of the comoving gauge curvature perturbation, usually denoted $\R$, 
since $\R$ is constant on large (super-Hubble) scales for adiabatic modes 
and hence its value at late times can be trivially related to the 
predictions of an inflationary model (see, \eg, \cite{ll00}).  However, 
we will see that it is 
simplest to perform the linear Sachs-Wolfe calculation in terms of the 
zero-shear gauge curvature perturbation, $\psi_\sig$, where we denote 
zero-shear gauge \perts\ by the subscript $\sig$.  For matter 
domination, it is simple to relate $\psi_\sig$ to $\R$ by performing a 
gauge transformation from zero-shear to comoving gauge; using 
\eqs(\ref{qgtrns}) and (\ref{psigtrns}) gives
\beq
\psi_\sig = -\fr{3}{5}\R.
\label{phirmd}
\eeq

   In terms of Fourier modes at comoving wavevector $\mathbf{k}$, on large 
scales $\R(\mathbf{k},t) = \Rpr(\mathbf{k})$, where $\Rpr(\mathbf{k})$ is 
the constant primordial value of $\R(\mathbf{k})$, \ie\ the value at early 
times sufficiently later than Hubble exit during inflation.  However, on 
small scales $\R$ decays during radiation domination.  The total decay 
incurred through to matter domination is described by a linear transfer 
function $T(k)$, defined by
\beq
\R(\mathbf{k},t_m) = T(k)\Rpr(\mathbf{k}),
\label{rtransfnc}
\eeq
where $k \equiv |\mathbf{k}|$ and $t_m$ is some time during matter 
domination.  ($T(k)$ will be 
distinguished from the temperature $T$ by the presence of its argument, 
$k$.)  The transfer function $T(k)$ approaches unity at small $k$, and decays 
roughly like $k^{-2}$ for $k \agt k_{\rm eq}$, where $k_{\rm eq}$ is the 
wavenumber which enters the Hubble radius at matter-radiation equality 
(see, \eg, \cite{ll00}).  $T(k)$ can be calculated accurately numerically, 
\eg\ using packages such as \textsc{camb} \cite{lcl00,notecamb}.  
Combining \eqs(\ref{phirmd}) and (\ref{rtransfnc}) gives
\beq
\psi_{\sig}(\mathbf{k},t_m) = -\fr{3}{5}T(k)\Rpr(\mathbf{k}).
\eeq

   The \pert\ $\psi_\sig$ is constant during matter domination, but it decays 
once a cosmological constant becomes important.  This decay is independent of 
scale and can be described by a function $g(t)$ via
\beq
\psi_{\sig}(\mathbf{k},t) \equiv g(t)\psi_{\sig}(\mathbf{k},t_m),
\label{gdefn}
\eeq
which gives
\beq
\psi_{\sig}(\mathbf{k},t) = -\fr{3}{5}g(t)T(k)\Rpr(\mathbf{k}).
\label{psiR}
\eeq
The function $g(t)$ approaches unity and zero at early and late times, 
respectively, and solving the linearized dynamical Einstein equation for 
$\psi_\sig$ gives
\beq
g(t) = \fr{5}{2}\fr{\Omega_{m,0} \Hb_0^2\Hb}{a} \int^t \fr{dt'}{a^2\Hb^2}.
\label{growth}
\eeq
Here $\Hb_0$ is the background Hubble rate today, and $\Om_{m,0}$ is the ratio 
of matter to total energy density today.  Using this last expression we 
can derive the useful relation
\beq
\fr{\dot g}{\Hb} + g = \Om_m\ld(\fr{5}{2} - \fr{3}{2}g\rd).
\label{greln}
\eeq

   The statistics of the assumed Gaussian random primordial fluctuations are 
completely described by the power spectrum $\PR(k)$, defined by
\beq
\bra\Rpr(\mathbf{k})\mathcal{R}^{\rm{pr}\ast}(\mathbf{k}')\ket
   = 2\pi^2\delta^3(\mathbf{k} - \mathbf{k}')\fr{\PR(k)}{k^3}.
\label{Rstat}
\eeq
With this definition $\PR(k)$ is dimensionless, and it is constant for a 
scale-invariant spectrum.

\section{Gauge invariance of anisotropies}
\label{ginvarapp}

   The calculated temperature anisotropy cannot depend on the coordinate 
choice (in the metric framework) or congruence choice (in the covariant 
framework), since the anisotropy is directly observable.  A 
number of workers have demonstrated this explicitly in the past in the 
metric formalism (see, \eg, \cite{sxek95,hn99}).  Nevertheless, it will 
be useful to demonstrate the result explicitly using the present covariant 
framework, as it will help to illuminate the issues involved.

   We wish to demonstrate explicitly that the expression 
\eq(\ref{templin}) for the anisotropy observed at $R$, namely
\beq
\fr{\del T(n^\mu)}{\bar{T}_R} = \int_{\bar{t}_R}^{\bar{t}_E} \del(H_nN)dt
   + \ld.\ld(\bar{H}\del t_D + n_\mu\del t_B^{;\mu}\rd)\rd|_R^E,
\label{templinapp}
\eeq
is invariant under arbitrary linear gauge transformations, if the point $R$ 
and the four-velocity of the observer are held constant.  Such a 
transformation changes the \hsfs\ of constant time according to
\beq
t \ra t - \del t,
\label{gtrans}
\eeq
for small temporal shift $\del t = \del t(x^\mu)$.  Equivalently, it 
changes the congruence orthogonal to the slicing by the spatial gradient 
of $\del t$,
\beq
u^\mu \ra u^\mu + D^\mu\del t,
\label{utrans}
\eeq
at linear order.  Under this change in $u^\mu$, each term in the integrand 
in \eq(\ref{templinapp}) will change, and the change in the slices 
$\Sig_{\bar{t}_E}$ and $\Sig_{\bar{t}_R}$ implied by \eq(\ref{gtrans}) 
means that the displacements $\del t_D$ and $\del t_B$ to the last 
scattering \hsf\ $\Sig_{\rm LS}$ and to the reception point $R$ will also 
change.

   Straightforward but lengthy calculations using the definitions 
\eq(\ref{acceldef}) 
to (\ref{sigdef}) give the following transformations under \eq(\ref{utrans}), 
to first order:
\bea
\theta &\ra& \theta + D^2\del t + 3\dot{\Hb}\del t,\label{thetatrans}\\
\sig_{\mu\nu}n^\mu n^\nu &\ra& 
   \sig_{\mu\nu}n^\mu n^\nu + n^\mu n^\nu D_\mu D_\nu\del t
    - \oo{3}D^2\del t,\\
a_\mu n^\mu &\ra& a_\mu n^\mu + n^\mu(\dot{\del t})_{;\mu},
\eea
where $D^2 \equiv D^\mu D_\mu$ is the physical Laplacian.  
Note that, by definition, after the gauge transformation (\ref{gtrans}) all 
quantities in the integrand in \eq(\ref{templinapp}) are to be evaluated 
at the new event temporally displaced by $\del t$ from the original event.  
This makes no difference at linear order to quantities that vanish at 
zeroth order (such as $\sig_{\mu\nu}$ and $a_\mu$), but accounts for the 
term $3\dot{\Hb}\del t$ in \eq(\ref{thetatrans}).  Next, considering that 
the quantity $Ndt$ is the proper time interval along $u^\mu$ between \hsfs\ 
separated by coordinate time interval $dt$, we can easily derive the 
linear transformation law for the lapse perturbation,
\beq
\del N \ra \del N + \dot{\del t}.
\label{lapsetrans}
\eeq
Combining \eqs(\ref{thetatrans}) to (\ref{lapsetrans}) with \eq(\ref{delhnn}), 
we have the linear transformation of the integrand of \eq(\ref{templinapp}),
\bea
\del(H_nN) &\ra& \del(H_nN) + (\Hb\del t)\dot{} + 
   n^\mu n^\nu D_\mu D_\nu\del t\nn\\
   && +\,n^\mu(\dot{\del t})_{;\mu}\\
   &=& \del(H_nN) + \fr{d}{dt}\ld.\ld(\Hb\del t + n_\mu\del t^{;\mu}\rd)
       \rd|_{v^\mu}.
\eea
This last line follows from the previous line by straightforward algebra, 
and contains a coordinate time derivative along the null geodesic $\O$.  
Therefore the integral in \eq(\ref{templinapp}) transforms according to
\beq
\int_{\bar{t}_R}^{\bar{t}_E} \del(H_nN)dt 
   \ra \int_{\bar{t}_R}^{\bar{t}_E} \del(H_nN)dt
     + \ld.\ld(\Hb\del t + n_\mu\del t^{;\mu}\rd)\rd|^E_R.
\label{inttrans}
\eeq

   Now all we need are the transformations for the boundary terms in the 
expression for the temperature anisotropy, \eq(\ref{templinapp}).  By 
\eq(\ref{gtrans}), the temporal displacement $\del t_{\rm disp}$ between 
any \hsf\ of constant $t$ and some fixed \hsf\ must transform like
\beq
\del t_{\rm disp} \ra \del t_{\rm disp} - \del t.
\eeq
Applying this expression to $\del t_{\rm disp} = \del t_D$ and 
$\del t_{\rm disp} = \del t_B$ gives for the transformation of the boundary 
terms
\bea
\ld.\ld(\Hb\del t_D + n_\mu\del t_B^{;\mu}\rd)\rd|^E_R 
   &\ra& \ld.\ld(\Hb\del t_D + n_\mu\del t_B^{;\mu}\rd)\rd|^E_R\nn\\
   &-& \ld.\ld(\Hb\del t + n_\mu\del t^{;\mu}\rd)\rd|^E_R.
\label{bdrytrans}
\eea
Combining \eqs(\ref{inttrans}) and (\ref{bdrytrans}) we finally find
\beq
\fr{\del T(n^\mu)}{\bar{T}_R} \ra \fr{\del T(n^\mu)}{\bar{T}_R},
\eeq
so that the temperature anisotropy is invariant under linear transformations 
of the congruence $u^\mu$, or equivalently the slicing $\Sig_t$, used in 
the calculation, if the point of observation and the four-velocity of the 
observer are held constant.  This of course was to be expected since the 
anisotropies are observable.

   Note that we are free to vary the ``background times'' $\bar{t}_R$ 
and $\bar{t}_E$ at first order.  Through \eq(\ref{tempbgnd}) this freedom 
simply shifts the temperature perturbation $\del T(n^\mu)$ by an irrelevant 
constant.  Indeed, we can use this freedom to fix $\del T(n^\mu)$ such that 
its mean over the whole sky vanishes for some particular observation point $R$,
\beq
\int \del T(n^\mu) d\Omega = 0,
\label{bgndtfreedom}
\eeq
so that $\bar{T}_R$ coincides with the mean temperature over the sky.  
Some authors consider it important to make this choice (see, \eg, 
\cite{dunsby97}).

\section{Gauge dependence of anisotropies}
\label{gdepapp}

   After demonstrating in Appendix~\ref{ginvarapp} the gauge {\em invariance} 
of the anisotropies described by \eq(\ref{templinapp}) for fixed observation 
point $R$ and observer four-velocity, we will now show how the anisotropies 
{\em do} depend on the gauge, when $R$ and the four-velocity are allowed 
to transform.  At linear order, we will see that such transformations will 
only effect the monopole and dipole anisotropies, as is well known.

   Consider again the gauge transformation
\beq
t \ra t - \del t,
\label{gtrans2}
\eeq
with corresponding change in the orthogonal congruence
\beq
u^\mu \ra u^\mu + D^\mu\del t.
\label{utrans2}
\eeq
Let us now evaluate the anisotropies using the general linear expression, 
\eq(\ref{templinapp}), but moving the reception point $R$ according to 
\eq(\ref{gtrans2}), and boosting the observer four-velocity according 
to \eq(\ref{utrans2}).  Since we move $R$, the corresponding emission 
point $E$ must also move.  If $R$ moves to the future, then the 
corresponding LSS will increase in diameter, and $E$ will move radially 
outwards.  We can schematically indicate the contributions to the change 
in the anisotropies under \eq(\ref{gtrans2}) by
\bea
\del\fr{\del T(n^\mu)}{\bar{T}_R}
   &=& \fr{\pd}{\pd E}\fr{\del T(n^\mu)}{\bar{T}_R} \del E
    + \fr{\pd}{\pd R}\fr{\del T(n^\mu)}{\bar{T}_R} \del R\nn\\
   && +\,\fr{\pd}{\pd u^\mu}\fr{\del T(n^\mu)}{\bar{T}_R} \del u^\mu.
\label{delttgtrn}
\eea

   The first term in \eq(\ref{delttgtrn}) arises due to the change in 
diameter of the LSS (and the entire past light cone).  As the LSS moves, it 
samples different perturbation modes, so the observed anisotropies 
change.  This effect is greatest for the structures at the smallest scales 
(comoving and angular), and was calculated in detail in \cite{zms07}.  
There it was shown that this contribution is of order
\beq
\fr{\pd}{\pd E}\fr{\del T(n^\mu)}{\bar{T}_R} \del E
   \sim \fr{\Hb(t_R)\del t}{\theta} \fr{\del T(n^\mu)}{\bar{T}_R},
\eeq
where $\theta$ is the angular scale of the feature in question.  Since 
$\del t$ and $\del T(n^\mu)$ are both first order quantities, this effect 
can be considered to be second order, and so will not be considered 
further here.  However, for large changes in observation time, substantial 
changes to the anisotropies will be observed \cite{zms07}.

   To calculate the second and third terms in \eq(\ref{delttgtrn}), 
note first that by displacing the observation point by $\del t$ and the 
observation four-velocity by $D^\mu\del t$, the displacements $\del t_D$ and 
$\del t_B$ at $R$ do not change under \eq(\ref{gtrans2}):
\bea
\del t_D(R) &\ra& \del t_D(R),\label{tdrtrans}\\
\del t_B(R) &\ra& \del t_B(R).
\eea
The displacements at the emission point $E$ still transform according 
to \eq(\ref{bdrytrans}),
\bea
\ld.\ld(\Hb\del t_D + n_\mu\del t_B^{;\mu}\rd)\rd|_E 
   &\ra& \ld.\ld(\Hb\del t_D + n_\mu\del t_B^{;\mu}\rd)\rd|_E\nn\\
   &-& \ld.\ld(\Hb\del t + n_\mu\del t^{;\mu}\rd)\rd|_E.
\label{ebdrytrans}
\eea
Combining \eqs(\ref{tdrtrans}) to (\ref{ebdrytrans}) with the 
transformation \eq(\ref{inttrans}) for the integral, we find that the 
anisotropies described by \eq(\ref{templinapp}) transform according to
\beq
\fr{\del T(n^\mu)}{\bar{T}_R} \ra \fr{\del T(n^\mu)}{\bar{T}_R} 
   - \ld.\ld(\Hb\del t + n_\mu\del t^{;\mu}\rd)\rd|_R.
\label{delttgtrn2}
\eeq

   To illuminate the nature of this change in the anisotropies, we can use 
the multipole expansion of the anisotropy, \eq(\ref{multipoleexp}).  If we 
align the polar axis along $D_\mu\del t$, we have
\bea
\Hb\del t &=& \sqrt{4\pi}\Hb\del t\,Y_{00}(n^\mu),\\
n_\mu\del t^{;\mu} &=& \sqrt{\fr{4\pi}{3}}|\del u^\mu| Y_{10}(n^\mu),
\eea
where $\del u^\mu \equiv D^\mu\del t$.  Combining these expressions with 
\eq(\ref{delttgtrn2}), the multipole expansion (\ref{multipoleexp}) gives
\bea
a_{00} &\ra& a_{00} - \sqrt{4\pi}\Hb\del t,\\
a_{10} &\ra& a_{10} - \sqrt{\fr{4\pi}{3}}|\del u^\mu|,\label{dipoleboost}
\eea
and all other multipoles are invariant under the transformation.  That 
is, only the monopole and dipole change.  Therefore the calculation of 
the higher multipoles is forgiving with respect to the care taken 
regarding gauge.  However, in Sec.~\ref{secmonodi}, where we calculate the 
dipole and monopole, we must be completely explicit about the specification 
of the frame in which we evaluate the dipole and the \hsf\ on which we 
evaluate the monopole.

   To close this discussion of gauge dependence, we note that going beyond 
first order, a boost at the observation point $R$ transfers power 
to {\em all} multipoles, and also distorts anisotropies through aberration 
\cite{cv02}.

\bibliography{bib}

\begin{thebibliography}{50}
\bibitem[{not({\natexlab{b}})}]{noterhoboost}
\bibinfo{note}{The energy density at some event, as seen by an observer with
  four-velocity $u^\mu$, will change at second order under a first order change
  of $u^\mu$.}
\bibitem[{not({\natexlab{c}})}]{notebgndtemp}
\bibinfo{note}{Temperature \perts\ {\em do} generally exist along an
  arbitrarily chosen background time slice which is a linear \pert\ from the
  physically defined $\Sls$, and this appears to be the source of the
  confusion.}
\bibitem[{not({\natexlab{d}})}]{noteextrcurv}
\bibinfo{note}{In the case that the congruence $u^\mu$ is \hsf\ orthogonal, the
  quantity $H_n$ is also simply related to the extrinsic curvature $K_{\mu\nu}$
  of the orthogonal slicing through $H_n = K_{\mu\nu}n^\mu n^\nu$.}
\bibitem[{not({\natexlab{e}})}]{notelensiso}
\bibinfo{note}{There is a subtlety here regarding the mapping of directions
  $n^\mu$ at $R$ to points $E(n^\mu)$ on the LSS. If gravitational lensing is
  important, there is the possibility that a single point on the LSS is mapped
  to multiple directions $n^\mu$ at $R$. Regardless, \eq(\ref{tnaffine}) can
  always be applied. A further subtlety here is the implicit assumption that
  the radiation released at $E$ is isotropic in the comoving frame. In
  principle, quadrupole and higher multipole anisotropies can exist at $E$, but
  our assumption of tight coupling in the plasma before last scattering implies
  that all such anisotropies are suppressed due to Thomson scattering (see,
  \eg, \cite{challinor04}).}
\bibitem[{not({\natexlab{f}})}]{notetwist}
\bibinfo{note}{This will not be possible in general if the congruence has
  non-vanishing twist $\om_{\mu\nu}$ \cite{wald84}. In this case a
  non-orthogonal slicing can be used, or a different orthogonal slicing could
  be chosen for each geodesic $\O$, since we only require the congruence and
  slicing to exist in the neighbourhood of each $\O$. At linear order, however,
  vanishing vector modes imply vanishing twist.}
\bibitem[{not({\natexlab{g}})}]{notecomove}
\bibinfo{note}{The construction of the comoving \hsfs\ in a thought experiment
  to attempt an estimate of the monopole variance would involve exchanging
  light signals to synchronize clocks, whereas uniform matter density slices
  can be defined by simple local measurements of $\rho$.}
\bibitem[{not({\natexlab{i}})}]{notemonomeas}
\bibinfo{note}{We can imagine an experiment that could in principle estimate
  $C_0$ by receiving information about the total and radiation energy densities
  on a slice inside the past light cone.}
\end{thebibliography}

\end{document}